\documentclass[twocolumn, tighten, times, twocolappendix]{aastex702}
\usepackage{amsmath}
\usepackage{subfig}
\usepackage{subcaption}
\usepackage{multirow}
\usepackage{booktabs}

\usepackage[dvipsnames]{xcolor} 

\shorttitle{Time-variable tertiary mass transfer in hierarchical triple systems \ \ }
\shortauthors{Rosselli-Calderon et al.}

\begin{document}
\title{Time-variable tertiary mass transfer in hierarchical triple systems}

\author[0000-0002-9537-1933]{Alejandra~Rosselli-Calderon}
\affiliation{Department of Astronomy and Astrophysics, University of California, Santa Cruz, CA 95064, USA}
\email{aleroca@ucsc.edu}

\author[0000-0002-1417-8024]{Morgan MacLeod}
\affiliation{Institute for Theory and Computation, Center for Astrophysics, Harvard \& Smithsonian, Cambridge, MA, 02138, USA}
\email{morgan.macleod@cfa.harvard.edu}

\author[0000-0003-2558-3102]{Enrico~Ramirez-Ruiz}
\affiliation{Department of Astronomy and Astrophysics, University of California, Santa Cruz, CA 95064, USA}
\email{enrico@ucolick.org}

\begin{abstract}
Most stars in the field are part of binary, triple, or higher-order systems, which we primarily observe in hierarchical configurations that allow long-term stability. Over long timescales, hierarchical triple systems interact primarily through gravitational dynamics, but more complex interactions arise when stellar evolution drives mass transfer. In this paper, we study tertiary mass transfer in hierarchical triple systems using three-dimensional hydrodynamic simulations. We find that the finite separation of the inner binary produces oscillations in the Roche potential that critically alter both the donor star’s variability and its mean mass-loss rate. For compact inner binaries, the mass-loss rate approaches that of a single accretor, while at larger separations the orbit-averaged mass-loss rate increases steeply, approximately as $\langle\dot M_d\rangle\propto \langle\dot M_{d,0}\rangle\left(1+C\  a_b^{6.1\pm0.9}\right)$. For the $\gamma = 5/3$ polytropic donor considered here, this scaling is consistent with the nonlinear response of Roche-lobe overflow to the leading quadrupolar perturbation of the inner-binary potential, which scales as $\delta\Phi_2\propto a_b^2$. We also examine the donor's response to periodic forcing from the inner binary and find a transition between individual and collaborative accretion that determines whether the transferred material can form and sustain a circumbinary disk. Our results show that the finite separation of the inner binary can strongly regulate both the rate and fate of tertiary mass transfer, constraining a poorly understood phase of hierarchical triple evolution.
\end{abstract}


\section{Introduction}

Hierarchical triple systems are a common outcome of star formation and play a central role in the evolution of compact-object binaries \citep{2014AJ....147...87T, 2023ASPC..534..275O}. These systems present physical processes that are unachievable with ordinary binary evolution, making them strong candidates for explaining myriad  astrophysical phenomena. Triple systems bring the study of gravitational dynamics together with stellar evolution over an enormous range of spatial and temporal scales. The coupling of these two leads to  outcomes that extend beyond single and binary evolution, such as low mass x-ray binaries \citep{low_mass_xray, V404_Cygni}, blue stragglers \citep{blue_stragglers, 2019ApJ...876L..33P, 2025ApJ...979L...1L}, tight binaries \citep{2013ApJ...766...64S, 2025A&A...704A.156R}, fast radio bursts \citep{2026ApJ..1000L..17S}, pinwheel outflows \citep{2024ApJ...966..163K}, and cataclysmic variables with high accretion rates \citep{2000ApJ...538..315M,2022MNRAS.514.1895K, 2025PASP..137g4201S}. They also present a variety of possible observational signatures, from type Ia supernovae \citep{2011ApJ...741...82T, 2023ApJ...950....9R} to gravitational wave emission \citep{2015PhRvD..92l4048L, 2017ApJ...834..200M,2020PhRvD.101j4053G}.
A triple body system can experience dynamical effects that can alter the eccentricity and inclination of the inner orbit through Lidov-Kozai cycles \citep{1962P&SS....9..719L, 1962AJ.....67..591K, 2016ARA&A..54..441N}, as well as tidal forces analogous to those that circularize binaries \citep{1998MNRAS.300..292K, Bataille2018}. It can also experience the forces induced by mass transfer episodes that can only be explained through stellar evolution. Some of these forces include the drag force from mass transfer \citep{Kummer2025}  and common envelope events \citep{triple_ce_outcomes, eccentric_ce} that can lead to mergers \citep{2019A&A...631A.126K, 2022PhRvD.106b3014S, 2025A&A...703A.123K} or tidal breakups that lead to three-body dynamically-chaotic interactions \citep{2012ApJ...760...99P, 2022ApJS..259...25H}. Mass transfer episodes in binaries and triples has the potential to leave imprints in the gravity modes and chemical composition of the donor star \citep{10.3389/fspas.2021.663026, 2026A&A...705A.164W}. Given the complexity to resolve such a large ranges in both length  and time scales, even the most basic questions, such as whether mass transfer remains stable, how angular momentum is redistributed, and which compact binaries survive, remain unanswered.

In recent years, the intricacies of this problem have been extensively explored, with some studies focusing on the gravitational  dynamics \citep{triple_roche_lobe_analytical, 2026ApJ..1007L..34K}, the stellar evolution and mass transfer episodes \citep{triple_roche_lobe, hydro_triple_ce,2021ApJ...907L..19V}, or the combination of them \citep{2022ApJS..259...25H}. The complexity of triple systems makes it hard to systematically approach these phases. Population synthesis studies have looked into bridging the gap and providing statics of the formation channels and final outcomes \citep{2022MNRAS.516.1406S, 2023ApJ...955L..14S, V404_Cygni, TRES} but are limited by the assumptions they make when the stars interact. Detailed studies of these interaction phases are crucial to inform these intricate stages in population synthesis codes. Further, the uncertainties in our understanding of the interactive phases of triple systems propagate directly into many of the most important problems in modern astrophysics. 

One such poorly-understood interaction phase is tertiary mass transfer, where the outermost star expands, initiating a mass transfer episode onto the tighter inner binary. The outcome of tertiary mass transfer determines whether hierarchical triples produce interacting binaries, common-envelope events, compact-object mergers, or luminous transients \citep{triple_roche_lobe, TRES, Leigh2020, 2021ApJ...907L..19V,2026A&A...713A.184B}. Consequently, our uncertainty in modeling tertiary mass transfer limits our ability to predict gravitational-wave populations, electromagnetic counterparts, and the evolutionary pathways leading to compact binaries.


In this paper we study stable tertiary mass transfer using three-dimensional hydrodynamic simulations of a polytropic donor transferring mass to an equal-mass inner binary. We vary the inner-binary separation to determine when the binary behaves effectively as a point-mass accretor and when its finite separation alters the Roche geometry, donor mass-loss rate, and subsequent evolution of the transferred material. We follow how the time-dependent Roche potential modifies the mass-transfer stream and donor mass-loss rate, and subsequently how the transferred material is redistributed around the inner binary. This allows us to connect the geometry of the time-dependent potential to both the rate and fate of tertiary mass transfer.

This paper is divided as follows. In Section \ref{sec:analytical} we set up the problem and provide the analytical framework that motivates the study of tertiary mass transfer episodes. In Section \ref{sec: methods} we present the numerical hydrodynamic framework and methods used to analyze stellar oscillations. In Section \ref{sec: results} we discuss the variability in the Roche potential and its effects on mass loss and stellar oscillations, as well as the effects of binary separation on the possible formation of circumbinary disks. Lastly, Section \ref{sec: conclusions} presents our summary and conclusions.

\section{System definition and characteristic scales}\label{sec:analytical}

To understand the evolution of mass transfer in a hierarchical triple system we must set the scene by taking a look at the governing physics. We take a step back to the well-known binary case to create an analytical framework of the interactions that dominate at different scales. We use the results from the binary exploration to inform our initial parameters and later interpret our results.

\subsection{System geometry and point-mass limit}

We consider a hierarchical triple system composed of an extended, giant donor star (tertiary; $M_d$) and an accreting inner binary composed of two more compact objects ($m_1$ and $m_2$), which may be stars or stellar remnants. By construction, the separation between the tertiary and the center of mass of the binary, $a_{t}$, is larger than the binary's semi-major axis, $a_{b}$, ensuring a dynamically-stable hierarchical arrangement.  

\begin{figure}
    \centering
    \includegraphics[width=0.7\linewidth]{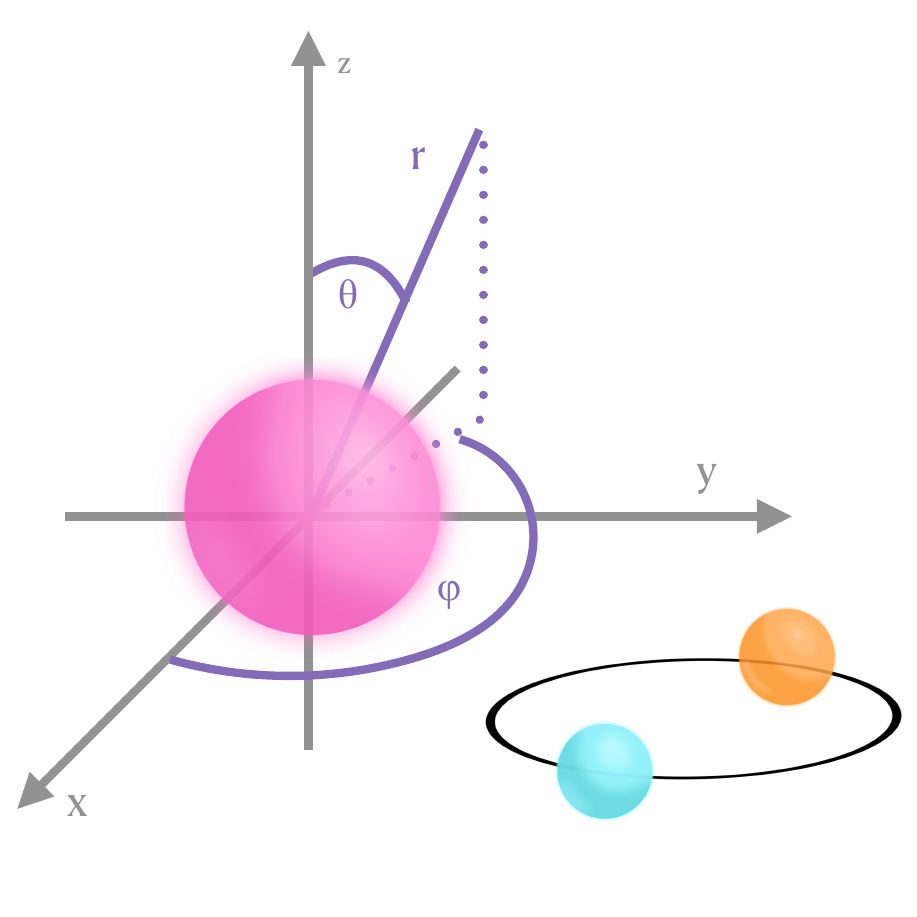}
    \includegraphics[width=1.0\linewidth]{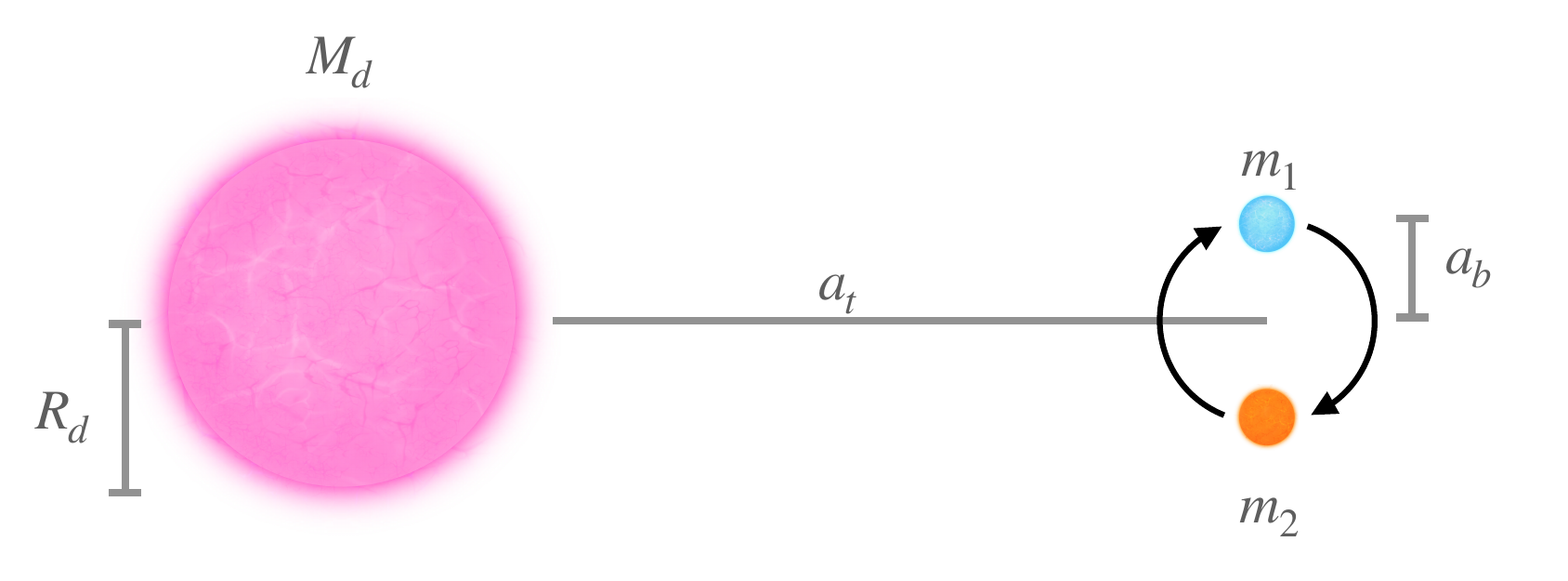}
    \caption{Schematic representation for the setup of our system of interest. \textit{Top:} Spatial representation of the numerical setup of the system. The tertiary polytropic star is placed at the center of the coordinate system with a point mass as a core. The inner binary is placed in the $x-y$ plane. Both the orbit of the inner binary and the orbit of the tertiary are initiated in the same $x-y$ plane. The simulations take place in a spherical-polar coordinate grid with $[r, \varphi, \theta].$
    \textit{Bottom:} Slice taken for $z=0$. The masses and length scales of the system are defined.}
    \label{fig:setup}
\end{figure}

The tertiary giant star which donates its mass to the binary has a total mass $M_{d} = M_{\rm{core}} + M_{\rm{env}}$ where $M_{\rm{core}}$ is the mass of the central core and $M_{\rm{env}}$ is the total mass of the gaseous envelope. The radius of the tertiary is $R_{d}$. The total mass of the binary is $m_{b} = m_1 + m_2$. We treat the stars in the binary to be equal masses, such that $m_1=m_2=0.5m_b$. The mass ratio of the system is $q=M_d/m_b$. A schematic of the setup of this system is presented in the bottom panel of Figure $\ref{fig:setup}$. 

There are some quantities that are worth defining when we look at the ordinary case of a single donor -- single accretor binary case. One of them is the total mass of the system, $M = M_{d} + m_b = M_{d} + m_1 + m_2$. This mass sets the outer orbital frequency of the binary around the tertiary to be $\Omega_{\rm{out}} = (GM/{a_{t}^3})^{1/2}$ and the orbital period to be $P_{\rm{out}} = 2\pi\Omega_{\rm{out}}^{-1}$. In comparison, the orbital frequency of the inner binary is $\Omega_{\rm{in}} = (Gm_{{b}}/a_{b}^3)^{1/2}$ with the corresponding orbital period  $P_{\rm{in}} = 2\pi\Omega_{\rm{in}}^{-1}$. 

By construction, we set the orbital plane of the inner binary to be the same as the orbital plane of the outer binary such that there is no relative inclination in the system. The orbits are also set to be circular, so the eccentricity of both the inner binary and the outer binary are zero. We make these choices in order to simplify the derivations that follow, but we note that these assumptions could be relaxed for systems with different eccentricities and inclinations. The orbital angular velocities of both orbits are set to be counterclockwise, with the axis of rotation pointing in the $+z-$direction. In the following subsections of Part $\ref{sec:analytical}$ we look at the different length scales of interest in the three-body hierarchical system problem. 

\subsection{Roche potential \& finding $L_1$}

We begin our study of length scales by using a simplified model system: a binary system. In the limit of $a_b \rightarrow 0$ we have a binary composed of a donor star of mass $M_{\rm{d}}$ and a companion of mass $m_b$ at a separation of $a_t$. For this system, we have a Roche potential

\begin{equation}
    \phi_{\rm{_R}} (\mathbf{r}) = - \frac{GM_d}{|\mathbf{r} - \mathbf{r}_d|}
    - \frac{Gm_b}{|\mathbf{r} - \mathbf{r}_b|} + \phi_{_\Omega}
\end{equation}

\noindent with 

\begin{equation}
    \phi_{_\Omega} = \frac{1}{2}\left[(\mathbf{\Omega}_{\rm{out}} \cdot \mathbf{r})^2
    - (\mathbf{\Omega}_{\rm{out}} \cdot \mathbf{\Omega}_{\rm{out}})
    (\mathbf{r} \cdot\mathbf{r})\right]
\end{equation}

\noindent where $\mathbf{r}$ is the radial distance to the center of the coordinate system, $\mathbf{r}_d$ is the distance from the coordinate origin to the center of mass of the donor and $\mathbf{r}_b$ is the distance to the point mass $m_b$. This potential leads the the classical Roche Lobe equipotentials solution for material that corotates with the system orbital frequency. From $\phi_{\rm{_R}}$ we can solve for five the stationary points in the co-rotating reference frame, $L_1-L_5$. We are most interested in the stationary point that resides between the donor and it companion, $L_1$. The distance between the donor star and $L_1$ can be approximated as the Roche radius (the radius of a sphere with equivalent volume to that of the Roche Lobe), 

\begin{equation}
    R_{\rm{L_1}} \approx a_t ( 0.5 + 0.227 \log_{10} q)
\label{eq: L_1 approx}
\end{equation}

\noindent with $q = M_{{d}}/m_b$ for the mass ratios of interest here \citep{1964BAICz..15..165P, 1976IAUS...73...85W}. 

Once the system is treated as a triple, the Roche potential then becomes 

\begin{equation}
    \phi_{\rm{_R}_3} (\mathbf{r}) = - \frac{GM_d}{|\mathbf{r} - \mathbf{r}_d|}
    - \frac{Gm_1}{|\mathbf{r} - \mathbf{r}_1|}
    - \frac{Gm_2}{|\mathbf{r} - \mathbf{r}_2|}
    + \phi_{_\Omega}
\end{equation}

\noindent where the indices $1,2$ denote the individual components of the inner binary. In this expression, we are still considering corotation with the outer orbital frequency $\Omega_{\rm out}$. 


    



For the triple case, we find the equipotentials and the stationary points numerically. Given that we have three terms in the potential, we expect to have the emergence of a sixth stationary point $L_6$. This stationary point will only arise when the three masses are not colinearly aligned or the inner binary is at a $90^{\circ}$ angle from the axis between its center of mass and the donor. We look at the variability in the Roche equipotentials for the case of a hierarchical triple arrangement and its effect on the distribution of gas in Section $\ref{sec: results}$.

\subsection{Radius of Circularization of Infalling Gas}

Whereas the Roche-potential calculation describes how the inner binary modifies the supply of material from the donor, the circularization and cavity scales describe the subsequent evolution of that material after it enters the Roche volume of the inner binary. In the limit of $a_b \rightarrow 0$, we now look at the behavior of the gas that leaves the Roche lobe of the donor and crosses into the Roche lobe of the accretor. Gas at the $L_1$ point will experience equal gravitational forces from the donor and the companion (that is, with the assumption that both of them are point masses). The gas at this point will have an angular momentum that cannot be dissipated. Unless the gas directly collides with the accretor, it will find a radius of stability where it will settle and form a rotationally-supported disk. At the $L_1$ point, the speed of sound of the gas is much less than the orbital velocity around the accretor, $c_s \ll v_{\rm{orb}}$. This allows us to write the equation of motion of a test particle as 

\begin{equation}
    \ddot{\tilde{\mathbf{r}}} + 2 \mathbf{\Omega} \times \dot{\tilde{\mathbf{r}}} = - \nabla {\phi_{\rm{_R}}}
\end{equation}

\noindent where $\tilde{\mathbf{r}}, \mathbf{\Omega}, {\phi_{\rm{_R}}}$ are the distance to the accretor, the angular velocity and the effective potential, respectively. 

Assuming no further torques after the gas crosses the $L_1$ point, we can set the specific angular momentum  of the gas at $L_1$, $l_{\rm{L_1}}$, equal to the Keplerian angular momentum of the gas at the radius of disk around the accretor, $l_k$. These two are given by

\begin{equation}
    l_{\rm{L_1}} 
    = \Omega_{\rm{out}}(a_t - R_{\rm{L_1}})^2
\end{equation}

\begin{equation}
    l_k = \sqrt{Gm_bR_c}
\end{equation}

\noindent with $R_c$ being the circularization radius at which gas will settle in a disk and $\Omega_{\rm{out}}$ the orbital frequency of the orbiting accretor (analogous to the outer orbital frequency of the center of mass of the inner binary around the donor for the case of a hierarchical triple). We use the approximation from Equation \eqref{eq: L_1 approx} to solve for $R_c$ in terms of $q$ and $a_t$ 

\begin{equation}
    \frac{R_c}{a_t} \approx (0.5 - 0.0986 \ln q)^4 (1+q).
\end{equation}

\noindent This $R_c$ is an approximation for  the distance away from $m_b$ at which a disk will form. Analytically, the material settles into a perfect ring at a fixed $R_c$, but in reality it will form a disk with a physical width and height. When a disk forms around a single accretor, the inner and outer radii of the disk depend on the mass ratio, mach number, viscosity, and temperature \citep{1981ARA&A..19..137P, 1997MNRAS.285..288L, Blondin_2000}. 

For the triple-system case where the accretor becomes the inner binary, the disk formation will additionally depend on the binary separation $a_b$. The additional angular momentum from the inner orbit has the ability to open up a cavity within the disk. We look at the radius of this cavity, $a_{\rm{cav}}$ and how it compares to the circularization radius in the next section.

\subsection{Radius of Circumbinary Disk Cavity }\label{sec:cavity}

The circularization radius of the infalling gas gives us an idea of the regime of inner binary separations to explore in the context of the formation of circumbinary disks. We now focus on the gas dynamics around the inner binary alone, neglecting the presence of the tertiary companion, to study circumbinary disks. Here we take $a_b > 0$ and explore the cavity that is cleared by the orbit of the binary accretor. 

We follow the calculations done in \cite{cavity_radius_1994} where they study the dynamics of a disk around a binary. They assume the disk to be geometrically thin, coplanar with the binary and governed by gas pressure and turbulent viscosity. We apply Fourier decomposition of the time-periodic gravitational potential of the binary in the inertial frame, $\Phi$, by separating the harmonic components of the disk. The perturbative potential is defined as

\begin{equation}
    \Phi (r, \varphi, t)
    =
    \sum_{m\ell} \phi_{m\ell}(r) \ 
    \text{exp}\left[ i(m\varphi - \ell {\Omega}_{\rm{in}} t)\right],
\end{equation}

\noindent where $m\geq0$, to avoid double counting modes, is the azimuthal harmonic number, $\ell$ is the time-harmonic number, $\phi_{m\ell}(r)$ is the radially variable potential component and ${\Omega}_{\rm{in}} = (Gm_b/a_b^3)^{1/2}$ is the mean motion of the binary. Each combination of $(m,\ell)$ gives rise to a set of Lindblad resonances, with $\phi_{m\ell}$ giving the respective value for the potential. From this potential, we conclude that the leading nonzero harmonic for an equal-mass, circular binary is $(m,\ell)=(2,2)$. Details of this calculation are presented in Appendix \ref{app: harmonics}. 

Solving for $(m,\ell)=(2,2)$ we get that the leading resonance occurs at $1.31 \ a_b$. Once \cite{cavity_radius_1994} carry out smooth-particle hydrodynamic simulations that account for the viscosity of the disk, they find that the cavity opened has a radius of $1.68 \ a_b$, which is reported in their Table 1. We take their numerically found value as a benchmark to estimate the cavity radius opened by the inner binary in our simulations.

\begin{table*}[ht]
\centering
\begin{tabular}{@{}lllllllllll@{}}
\toprule
$a_t$ & $L_{1,t}$ & $L_{1,b}$ & $a_{\rm circ}$ & $a_b$ & $a_{t,\rm near}$ & $a_{t,\rm far}$ & $L_{1,\rm overlap}$ & $L_{1,\rm separated}$ & $\Delta L_1$ & $a_{\rm cav}$ \\
\midrule
      &           &           &               & 1.0  & 3.1   & 4.1   & 1.3353  & 1.4366 & 0.1013 & 1.68 \\
3.6   & 1.4014     & 2.1986     & 0.65          & 0.6  & 3.3   & 3.9   & 1.3768  & 1.4139 & 0.0371 & 1.01 \\
      &           &           &               & 0.3  & 3.45  & 3.75  & 1.3951 & 1.4044 & 0.0093 & 0.504 \\
      &           &           &               & 0.15 & 3.525 & 3.675 & 1.3997 & 1.4021 & 0.0024 & 0.252 \\
\bottomrule
\hline
\end{tabular}
\caption{Here we present some of the length scales of interest as analytically calculated. The ten columns present the separation of the tertiary $a_t$, the approximate distance of the $L_1$ point from the tertiary (when calculated for a single accretor), the approximate distance of the $L_1$ point from the center of mass of the accretor (when calculated for a single accretor), the separation of the binary $a_b$, the point of nearest approach of one of the components of the inner binary to the tertiary $a_{t, \rm{near}}$, the point of farthest distance of one on the components of the inner binary to the tertiary $a_{t, \rm{far}}$, the location of the instantaneous stability point (measured as a distance from the tertiary) analogous to $L_1$ when the lobes are at the largest overlap and largest separation, $L_{1, \rm{overlap}}$ and $L_{1, \rm{separated}}$, the difference between these two values, $\Delta L_1$, and the radial distance from the center of mass of the inner binary to the inner edge of the disk cavity opened, $a_{\rm{cav}}$. The values in the first four columns remain the same for the whole suite of simulations.
}
\label{table: lengthscales}
\end{table*}

\subsection{Point-Mass and Perturbative Limits}

An important limit occurs when the separation of the inner binary, $a_b$, is small compared to the distance $r$ from the center of mass of the inner binary to the point where its gravitational potential is evaluated. Defining $\epsilon=a_b/r\ll1$, the gravitational potential of the inner binary can be expanded about its center of mass. The leading term is the monopole potential of a single object of mass $m_b=m_1+m_2$. For the equal-mass binary considered here, the odd multipoles vanish by symmetry, leaving the quadrupole as the leading finite-separation correction,

\begin{equation}
    \Phi_b = \Phi_0 + \delta\Phi_2+ \mathcal{O}\left( \epsilon^4 \right),
\end{equation}

\noindent with

\begin{equation}
    \frac{\delta\Phi_2}{\Phi_0} \propto  \left( \frac{a_b}{r}\right)^2.
\end{equation}

Thus, as $a_b\rightarrow0$, the tertiary recovers the ordinary binary point-mass Roche geometry, while the leading finite-separation perturbation scales approximately as $a_b^2$. 
Roche-lobe overflow is sensitive to the small potential difference between the donor surface and the mass-transfer region. Even when $|\delta\Phi_2|/|\Phi_0|\ll1$, the quadrupolar perturbation can be comparable to the effective Roche-lobe overfill and substantially alter the mass-loss rate. This provides the basis for interpreting the separation dependence of the donor mass-loss rate discussed in Section \ref{sec:oscillations}.

We calculate all of the values presented in Section \ref{sec:analytical} for inner binary separations $a_b = [0.15, 0.3, 0.6, 1.0]$ which correspond to the values for the full set of simulations in this study. We present these values in Table \ref{table: lengthscales} to have as reference for the analysis of the results and comparisons to simpler two-body cases. We present the conditions for the simulations in the following section. 

\section{Methods and Numerical Setup} \label{sec: methods}

In this work, we focus on the mass transfer of a tertiary star onto a nearby orbiting binary. Due to the nature of this problem, we necessitate a global three-dimensional setup.  We set up the simulations in three dimensional spherical polar coordinates, while on the non-rotating reference frame of the tertiary. The top diagram in Figure \ref{fig:setup} illustrates the overall three dimensional setup. Given the difference in size scale between the compact objects and the giant, we represent the two objects in the binary as point masses.  Similarly, the core of the tertiary is represented by a point mass enclosed by an extended gaseous envelope.

We adopt a similar setup to that in \cite{Morgan_2018} in a non-rotating reference frame with the tertiary at the origin and the binary orbiting about itself and about the tertiary. The orbit of the inner binary is prograde in reference to the outer orbit. We tested both rotating and non-rotating coordinate frames, and we chose the non-rotating frame for the calculations in this work because we found that it preserved the hydrostatic equilibrium of polytropic stars more consistently. In this section we present our numerical setup and initial conditions. 

\subsection{Units}

In order to keep the simulations scalable to a wide range of astrophysical systems, we utilize code units where $G = M_d =R_d = 1$. By fixing these, we can derive the units for velocity as $({GM_dR_d^{-1}})^{1/2} = 1$. The appropriate units are for the rest of the quantities of interest are: $({G^{-1}M_d^{-1}R_d^3})^{1/2}$ for time, $M_dR_d^{-3}$ for  density, $GM_d^2R_d^{-4}$ for  pressure, $({GM_d^3R_d^{-3}})^{1/2}$ for  mass transfer rate, and $({GM_d^3R_d})^{1/2}$ for  angular momentum. 

\subsection{Gas equations}

We set up our problem in \texttt{Athena++}, an Eulerian grid-based code for astrophysical magnetohydrodynamics \citep{2020ApJS..249....4S}. \texttt{Athena++} numerically solves the following equations of compressible, adiabatic, inviscid, ideal gas dynamics: 

\begin{subequations}
\begin{align}
 \partial_T \rho + \nabla \cdot(\rho \mathbf{v}) &= 0, \\
 \partial_T(\rho\mathbf{v}) + \nabla\cdot(\rho \mathbf{v} \mathbf{v} + P \mathbf{I}) &=  \rho \boldsymbol{\alpha}_{\rm{ext}},  \\
 \partial_T E + \nabla \cdot [(E+P)\mathbf{v}] &= \rho\boldsymbol{\alpha}_{\rm{ext}} \cdot \mathbf{v},
\end{align}
\label{eq: gas}
\end{subequations}

\noindent where the first equation represents mass continuity, the second equation the evolution of momentum and the third equation the evolution of energy. In this set of equations, $\rho$ is the mass density, $\rho\mathbf{v}$ is the momentum density, and $E$ is the total energy density. The subscript $T$ in the partial derivatives denotes differentiation with respect to time. We define the total energy density $E=\varepsilon+\rho\mathbf{v}\cdot\mathbf{v}/2$, with $\varepsilon$ representing the internal energy density of the gas. We also have $P$ as the pressure, $\mathbf{I}$ as the identity tensor, and $\boldsymbol{\alpha}_{\rm{ext}}$ an an external acceleration source term. We note that the external acceleration is negative, making the right hand side of the last two equations above positive. We adopt an ideal gas equation of state, $P=(\gamma-1) \varepsilon$, where $\gamma$ is the adiabatic gas index. 

\subsection{Source terms}

The terms on the right-hand side of Equations $\ref{eq: gas}$ account for the external forces that act directly on the gas in our system. By construction, we add forces to the system by introducing the three bodies in the triple system and by taking into account self-gravitation for the gas. We can decompose $\boldsymbol{\alpha}_{\rm{ext}}$ into these components as

\begin{equation}\label{eq: external_acc}
    \boldsymbol{\alpha}_{\rm{ext}} = \boldsymbol{\alpha}_t + \boldsymbol{\alpha}_1 + \boldsymbol{\alpha}_2 + \boldsymbol{\alpha}_{\rm{sg}},
\end{equation}

\noindent where $\boldsymbol{\alpha}_t$, $ \boldsymbol{\alpha}_1$, and $ \boldsymbol{\alpha}_2$ are the gravitational accelerations associated with the core of the tertiary, accretor 1 and accretor 2 respectively (not to be confused with angular acceleration). The gravitational accelerations are negative, as they point from the gas to the gravitational source. The acceleration denoted by $\boldsymbol{\alpha}_{\rm{sg}}$ is that of the self-gravity of the gas. 

The gravitational acceleration of the component introduced by the tertiary's core takes the form of

\begin{equation}
    \boldsymbol{\alpha}_t = - \frac{GM_{\rm{core}}}{|\mathbf{r}|^3}\mathbf{r}
\end{equation}

\noindent where $\mathbf{r}$ is the distance vector from the origin of the coordinate system, due to the fact that the tertiary is placed at the origin. For the two point masses in the binary, the gravitational acceleration takes the following form

\begin{equation}
\begin{split}
        \boldsymbol{\alpha}_i &= - \frac{Gm_i}{|\mathbf{r}-\mathbf{r}_i|^3} (\mathbf{r} - \mathbf{r}_i) \\
        & \approx - Gm_i (\mathbf{r} - \mathbf{r}_i) f_{\rm{spline}} (|\mathbf{r} - \mathbf{r}_i|, r_{\rm{soft}})
\end{split}
\end{equation}

\noindent where the index \textit{i} are $1, 2$ to represent $m_1$ and $m_2$. The second row presents the numerically approximated value used to calculate the gravitational acceleration at small distances $r \lesssim 2r_{\rm{soft}}$ from the point particles. The spline function is defined as

\begin{flalign}
    & f_{\rm{spline}} (r, r_{\rm{soft}}) = \nonumber \\ &
    \begin{cases}
        r_{\rm{soft}}^{-3} \left[ \frac{4}{3} - \frac{6}{5}u^2 + \frac{1}{2} u^3\right] & 0 \leq u < 1 \\
        r^{-3} \left[ - \frac{1}{15} + \frac{8}{3}u^3 - 3u^4 + \frac{6}{5}u^5 -\frac{1}{6} u^6 \right] & 1 \leq u < 2 \\
        r^{-3} & u\geq 2
    \end{cases}
\end{flalign}

\noindent where $u \equiv r/r_{\rm{soft}}$ as defined in Equation A2 in \cite{1989Hernquist}. The use of a spline function is indispensable to treat gravity in a grid-based code. Near the point masses, $\mathbf{r}- \mathbf{r}_i \rightarrow 0$. In order to deal with this singularity, we introduced a softening radius $r_{\rm{soft}}$. Within a distance $|\mathbf{r} -\mathbf{r}_i| < 2r_{\rm{soft}}$ we can approximate gravity without encountering a singularity. This treatment also ensures that outside of $2 r_{\rm{soft}}$ we use the exact solution for Newtonian gravity. 

Lastly, the rightmost term in Equation \ref{eq: external_acc} is the acceleration due to the self-gravity of the gas. We treat this acceleration in an approximate way by using a ``static self-gravity" approach that only takes into account the enclosed mass measure from the original hydrostatic equilibrium profile, rather than by calculating it at every timestep. This takes the form of

\begin{equation}
    \boldsymbol{\alpha}_{\rm{sg}} \approx
    - \frac{G[M_{\text{enc},d}(r)]-M_d}{|\mathbf{r}|^3}
    \mathbf{r}
 \end{equation}

 \noindent where $M_{\text{enc},d}(r)$ is the static enclosed mass at a radius $r$. The choice of taking the approximate enclosed mass has the advantage of being significantly more computationally efficient at the cost of not accounting for any mass loss that takes place over the course of the simulation. Appendix A in \cite{Morgan_2018} explains the implications of this approximation on the conservation of angular momentum of the system. {We explore systems in which the fraction of mass loss from the system is $\lesssim 10^{-3}M_d$, thus the approximate treatment of self-gravity is sufficient for our purposes.}

 We now transition to looking at the external accelerations on the tertiary. Relative to the inertial frame, the acceleration of the tertiary's core is

 \begin{equation}
     \boldsymbol{\alpha}_{t, \rm{ext}} = 
     \frac{Gm_1}{|\mathbf{r}_1|^3} \mathbf{r}_1 +
     \frac{Gm_2}{|\mathbf{r}_2|^3} \mathbf{r}_2 +
     \boldsymbol{\alpha}_{t,\text{gas}}
 \end{equation}

 \noindent where the first two terms account from the gravitational acceleration due to the two components of the inner binary and the rightmost term is the acceleration due to the gas, defined as

 \begin{equation}
     \boldsymbol{\alpha}_{t,\text{gas}} = 
     \int_{\rm{vol}} 
     \frac{G \ \rho \ dV}{|\mathbf{r}|^3} \mathbf{r}.
 \end{equation}

\noindent We numerically compute the above integral for every timestep in the simulation.

\begin{equation}
    \boldsymbol{\alpha}_{i,\rm{ext}} =
    \frac{- GM_{\rm{core}}}{|\mathbf{r}_i|^3} \mathbf{r}_i +
    \frac{Gm_j}{|\mathbf{r}_i - \mathbf{r}_j|^3}
    (\mathbf{r}_i - \mathbf{r}_j) + \boldsymbol{\alpha}_{i, \rm{gas}}
\end{equation}

\noindent with the index $j \neq i$. The two point particles are also experiencing the force from the surrounding gas. We calculate that force as 

 \begin{equation}
     \boldsymbol{\alpha}_{i,\text{gas}} = 
     \int_{\rm{vol}} 
     \frac{G \ \rho \ dV}{|\mathbf{\tilde{\mathbf{r}}}_i|^3} \tilde{\mathbf{r}}_i.
 \end{equation}

\noindent Here, $\mathbf{\tilde{r}}_i$ is the distance away from the particle $m_i$ to the parcel gas.

\subsection{Parametrized cooling}\label{sec:cool}

In the interest of allowing the gas to settle into disk-like structures, as might happen in an astrophysical system, we introduce cooling to the system. We prescribe a cooling function for the gas similar to that implemented in \cite{Gammie2001}. We introduce a cooling timescale $t_{\rm{cool}}$ that depends solely on the density of the material. We define this timescale as

\begin{equation}
    t_{\rm{cool}} = (t_{\rm{max}}-t_{\rm{min}}) 
    \tanh \left[ \left( \frac{\rho}{\rho_c}
    \right)^{\xi} \right] + t_{\rm{min}},
\end{equation}

\noindent where $\xi$ is a parameter that sets the steepness of the slope between the minimum value of the timescale, $t_{\rm{min}}$, and the maximum value of the timescale, $t_{\rm{max}}$. Here, $\rho$ is the density of the material at each cell and $\rho_c$ is a critical density, such that material with densities higher than $\rho_c$ will not be cooled. 

We look at the equilibrium pressure of the gas, $P_{\rm{eq}}$, defined for an ideal gas as

\begin{equation}
    P_{\rm{eq}} =\frac{\rho}{\gamma} \ c^2_{s, \rm{eq}}
\end{equation}

\noindent where $\gamma$ is the adiabatic index of the gas and $c_{\rm{s,eq}}$ is its equilibrium sound speed. We compare $c_{\rm{s,eq}}$ to the Keplerian speed of the gas, 

\begin{equation}
    c_{s, \rm{eq}} = h \ v_k,
\end{equation}

\noindent with $h$ being a free parameter set to model the the aspect ratio of the disk, $h = H/R_{\rm{disk}}$, where $H$ is the height of the disk and $R_{\rm{disk}}$ is its cylindrical radius. In this formalism, $v_k$ is the Keplerian velocity of the gas, which we approximate as, 

\begin{equation}
    v_k =  \sqrt{G\ r_{\rm{com}} \left( \frac{m_1}{r_1^2} + \frac{m_2}{r_2^2}\right)}.
\end{equation}

\noindent For the Keplerian velocity calculation, $r_{\rm{com}}$ is the distance of the gas parcel to the center of mass of the binary system comprised of $m_1$ and $m_2$, $r_1$ is the distance of this parcel to $m_1$ and $r_2$ is the distance to $m_2$. Here, we relate the centripetal acceleration of a parcel of gas in orbit to the gravitational acceleration induced by the two components of the binary. We update the pressure of each parcel of gas as

\begin{equation}
    \Delta P = - (P_i - P_{\rm{eq}}) \left( 1 - e^{-\Delta t/t_{\rm{cool}}}\right),
\end{equation}

\noindent where $P_i$ is the initial pressure and $\Delta P$ and $\Delta t$ are the discretised pressure and time steps respectively. In order to make sure that we are not adding heat into the system, we limit $\Delta P <0$ values when we advance in time. Additionally, we add a condition where we only cool in regions where $r > R_d + \Delta r$ with $\Delta r = 0.1$ to prevent material from the star (and the region closest to it) from cooling down.

A sweet spot would cool material from the mass transfer stream and around of the inner binary without dramatically changing the structure of the donors and the mass loss rate. The parameter exploration comes into play in the definition of $t_{\rm{cool}}$, where $t_{\rm{min}}$, $t_{\rm{max}}$ and $\rho_c$ will determine where the cooling takes place and at what rate. We make the choice of $t_{\rm{min}} = 0.1$, setting a base value that will dictate the fastest cooling rate. We found that smaller values for the cooling timescale yield changes in pressure that destabilize the system. In order to prevent material from cooling, we set $t_{\rm{max}} = 10^{10}$, which gives a timescale longer than the simulation time, so the material effectively does not cool. We explore different values for $\rho_c$. For small values of $\rho_c$, only very diffuse material cools down, so the majority of the mass being transfer is unaffected by the cooling function. For large values of $\rho_c$, the material inside of the polytropic star will be cooled.  In Appendix \ref{app:cooling} we present our study for values of $\log \rho_c = [-3, -4, -5]$ and compare the change in mass of the donor for the case with no cooling, $\rho_c = -\infty$. We decide to run all of the simulations in this work with $\rho_c = 10^{-4}$. The scale height was set to $h=0.2$, corresponding to a thick disk \citep{M_sta_2019}, and the value for $\xi$ was arbitrarily set to $10^{10}$ to make $t_{\rm{cool}}$ effectively a step function.

\subsection{Initial Conditions}

We begin by discussing the initial conditions of the donor. We initialize the tertiary donor at the center of the coordinate system. The tertiary constitutes of a core represented by a particle of mass $M_{\rm{core}} = 0.5$ and radius $R_{\rm{core}} = 0.1$ surrounded by an envelope of mass $M_{\rm{env}} = 0.5$ and radius $R_d - R_{\rm{core}} = 0.9$ for a total mass of the donor $M_d = 1.0$ and total radius $R_d = 1.0$ all in code units. We set the rotation of the envelope to be synchronous with the motion of the outer orbit, such that $\Omega_{\rm{env}} = \Omega_{\rm{out}}$. This gaseous envelope adopts a polytropic profile with equal polytropic and adiabatic indexes $\Gamma_s = \gamma = 5/3$. We map a 1D hydrostatic profile which is presented in Appendix \ref{app:convergence}. Since we are using a grid-based hydrodynamics code, we cannot embed this hydrostatic equilibrium profile directly onto an absolute numerical vacuum. In order for the profile to remain in equilibrium, we must add a transitional profile from the ``edge'' of the star at a pressure of $10^{-6}$ to the background ``vacuum'' at pressure of $10^{-12}$. A polytropic envelope has a structure given by

\begin{subequations}
\begin{eqnarray}
\frac{dP}{dr} =& -g \rho, \\
\frac{d\rho}{dr} =& \frac{-g\rho^2}{\Gamma_s P},\\
\frac{dm}{dr} =& 4\pi r^2 \rho,
\end{eqnarray}
\end{subequations}

\noindent where $g$ is given by $g=GM(r)/r^2$ and $M(r)$ is the mass enclosed of the donor at a radius $r$. Outside of the polytrope's radius of $R_d$ we place a low density ``corona'' to remain in hydrostatic equilibrium as we transition from the ``edge'' of the star to the background ``vacuum''. The structure of this ``corona'' is given by

\begin{equation}
    \frac{d\rho}{dr} = -\frac{g\rho\gamma}{c^2_{s,c}},
\end{equation}

\noindent where the sound speed $c_{s,c}$ is constant. The envelope's velocity in $r$ and $\theta$ are initialized at zero.

We initiate each one of the components of the inner binary as a point particle of mass $m_1 = m_2 = 1.5 M_d$ such that $q=1/3$. The center of mass of the inner binary is placed at the location  $r =a_t = 3.6R_d$, $\theta = \pi/2$, $\varphi = 0$, equivalent to $x=3.6R_d, \ y=0, \ z=0$. The inner binary point particles are placed at $a_t \pm a_b/2$ along the $x-$axis. We explore the cases where $a_b = [0.15, 0.3, 0.6, 1.0]$. Both the inner and outer orbits are given angular momentum in the $+z-$direction. The inclinations of the orbits are zero, which keeps the system coplanar in the $x-y$ plane. The top panel of Figure \ref{fig:setup} illustrates the three-dimensional setup of the system in the simulation grid, while the bottom panel of Figure \ref{fig:setup} shows a slice of the setup at the plane where $z=0$. It outlines the masses and length scales presented above. 


\subsection{Mesh and domain}

We set up the simulations on a spherical polar mesh centered at the core of the tertiary star. The grid has uniformly spaced zones in the $\theta$ and $\varphi$ coordinates and logarithmically spaced in the $r$ coordinate. This allows the shape of the zones to remain roughly cubic. We set up the mesh with $16$ blocks in the $r$ coordinate, $16$ blocks in the $\theta$ coordinate and $24$ blocks in the $\varphi$ coordinate.{ Overall, our simulation is divided into $144$ of zones in $r$, $96$ zones in $\theta$ and $192$ zones in $\varphi$, for a total of $2.6 \times 10^{6}$ zones.} The total simulation domain goes from $[0.3, 30]$ in the $r$ coordinate, $[0, \pi)$ in $\theta$ and $[0,2\pi)$ in the $\varphi$ coordinate.  The outer boundary in the $r-$direction has a ``diode'' condition which allows for the outflow of material but does not allow inflow. The inner $r$ boundary is ``reflective" so no material can go beyond it. In the $\theta$ coordinate, we impose a ``polar" boundary that allows for material to flow freely through the poles. For the $\varphi$ coordinate, the boundary conditions are periodic so that material can wrap around the domain at the $\varphi = 0 = 2\pi$ border. 


We additionally implement two levels of nested adaptive mesh refinement (AMR). In AMR, the refinement of a zone of interest increases by a factor of two for each level. In contrast to static mesh refinement, in AMR the zone of interest is allowed to shift over the course of a simulation. This has the advantage of finely resolving a small dynamic zone rather than a large static chunk of the domain, but the disadvantage of losing information when zones are refined and de-refined. 

We center the AMR levels at the center of mass of the inner binary. Since the simulations are performed in the inertial reference frame of the tertiary, the center of mass of the inner binary has a translational orbit around the center of the coordinate system. The highest refinement level goes from $1.1 \leq r < 5$, and $-0.5 < \Delta \theta \leq 0.5$ and $-0.5 < \Delta \varphi < 0.5$ radians (roughly $30^{\circ}$) away from the nearest of the components of the inner binary. This ensures that the whole inner binary remains within this refinement level at all times. The next refinement level goes from $1.43 \leq r \le 6.5$, $-1.0 < \Delta \theta \leq -0.5$ and $0.5 < \Delta \theta \leq 1.0$, and $-1.0 < \Delta \varphi < -0.5$ and  $0.5 < \Delta \varphi < 1.0$(roughly from $30^{\circ}$ to $60^{\circ}$) away from the nearest of the components of the inner binary. We force the system to not de-refine for values of $r<1.1$ to make sure the polytropic envelope has a constant resolution spatially and over time throughout. As the binary translationally orbits, the angular size of the refinement levels in the $\varphi$ coordinate variates to include the totality of the mesh block that contains a zone that falls within the refinement condition. 

A complexity of using spherical polar orbits is that singularities arise at $\theta = 0$ and $\theta= \pi$.  The zones around these poles become very narrow in the $\varphi$ direction, which consequently make the timestep increasingly short. To get past this computational hurdle, we create effectively larger zones by averaging conserved quantities over neighboring zones at the poles. 

When placing a hydrostatic profile in a three-dimensional grid, we encounter the main limitation of grid-based codes: mapping a continuous profile to a set of discretised blocks. The theoretical hydrostatic equilibrium envelope needs to be relaxed in order to find its numerical hydrostatic equilibrium. This limitation improves as we go to higher resolutions with smaller blocks and more zones. The downside with higher resolution grids is that they are significantly more computationally expensive. We test convergence of the hydrostatic profile by testing different resolutions, first in the $r-$coordinate and then in $\theta$ and $\varphi$. We present the details of this study in Appendix \ref{app:convergence}. Figure A\ref{fig:1d_convergence} shows the hydrostatic profiles after a time $t=50$ in simulation units. We can appreciate that for higher resolutions (smaller $dr/r$) the profiles track both the edge of the star and the transitional corona quite closely.

\subsection{Relaxation Scheme}

Another tool we employ when going from analytical to numerical hydrostatic equilibrium is damping the velocities of the gas. This lets the profile find a state of equilibrium by itself before the introduction of gravitational force from the companion binary accretors. When going from 1D to 3D, angular velocities arise. These have the potential of disturbing the envelope's structure. These small perturbations quickly lead to conditions that are out of equilibrium, such as shocks that completely unbind the gas. We begin the simulations by introducing a damping source term, 

\begin{equation}
    \boldsymbol{\alpha}_{\rm{damp}} = - \frac{\mathbf{v}}{\tau_{\rm{damp}}}.
\end{equation}

\noindent Here $\tau_{\rm{damp}}$ is the damping timescale. We also make a choice of a relaxation time $t_{\rm{relax}}$. The damping will only take place for times $t \leq t_{\rm{relax}}$ and in the $r-$ and $\theta-$ directions. To implement the damping, we set the initial value of the timescale $\tau_{\rm{damp,0}}=1$ for times $t<0.2 \ t_{\rm{relax}}$. We slowly increase the damping as a function of

\begin{equation}
    \tau_{\rm{damp}} = \tau_{\rm{damp,0}} \ 10^{2(t-0.2t_{\rm{relax}})/0.8t_{\rm{relax}}}.
\end{equation}
Therefore, by the time $t=t_{\rm{relax}}$, $\tau_{\rm{damp}}=100\tau_{\rm{damp,0}}$. 

The whole time while $t<t_{\rm{relax}}$, we have kept the polytropic profile in isolation. The two gravitational point masses from the inner binary were ``turned off'' meaning that their source terms are set to zero. We slowly introduced the source terms coming from the accretors by linearly increasing a multiplicative factor from 0 to 1. We do this for a time $t_{\rm{mass\_on}}$ that starts immediately after $t_{\rm{relax}}$, such that the source terms take full effect at a time $t_{\rm{relax}} + t_{\rm{mass\_on}}$. For our models, we find an optimized setup with $t_{\rm{relax}} = 15$ and $t_{\rm{mass\_on}} =10$. This means that the simulations are fully active by 25 dynamical times of the tertiary star.

\subsection{Stellar Oscillation Analysis}

In order to analyze the time-variable oscillations of the system, we need to isolate the harmonic oscillation of the star itself from those oscillations that are excited by the companion inner binary. Made for the field of astroseismology, the stellar oscillation code \texttt{GYRE} is the ideal tool to calculate the oscillations of the tertiary star \cite{gyre}. We input the properties of the star (mass $= 1.0$, radius $=1.0$, rotation of the envelope being corrotational with the outer orbit) and are able to obtain the frequencies of the harmonic oscillation along with their corresponding harmonic degree $\ell$, azimuthal order $m$ and radial order $n_r$.

\section{Results}\label{sec: results}

In this section, we present the results of our numerical simulations. We analyze the effects of the variability of the potential in the structure and distribution of mass, pressure, mach number and specific entropy of the gas. We then measure the mass loss of the donor in the context of overlapping Roche lobes. We look at how this overlap affects the oscillations of the tertiary. We focus on the structure near the binary accretors to understand the dynamics that allow for circumbinary disk formation. Lastly, we zoom out to view the overall picture of the pulsed mass ejection, noting the possible formation of spiral-arm-like structures. 

\begin{figure*}[h!]
    \centering
    \includegraphics[width=0.9\linewidth]{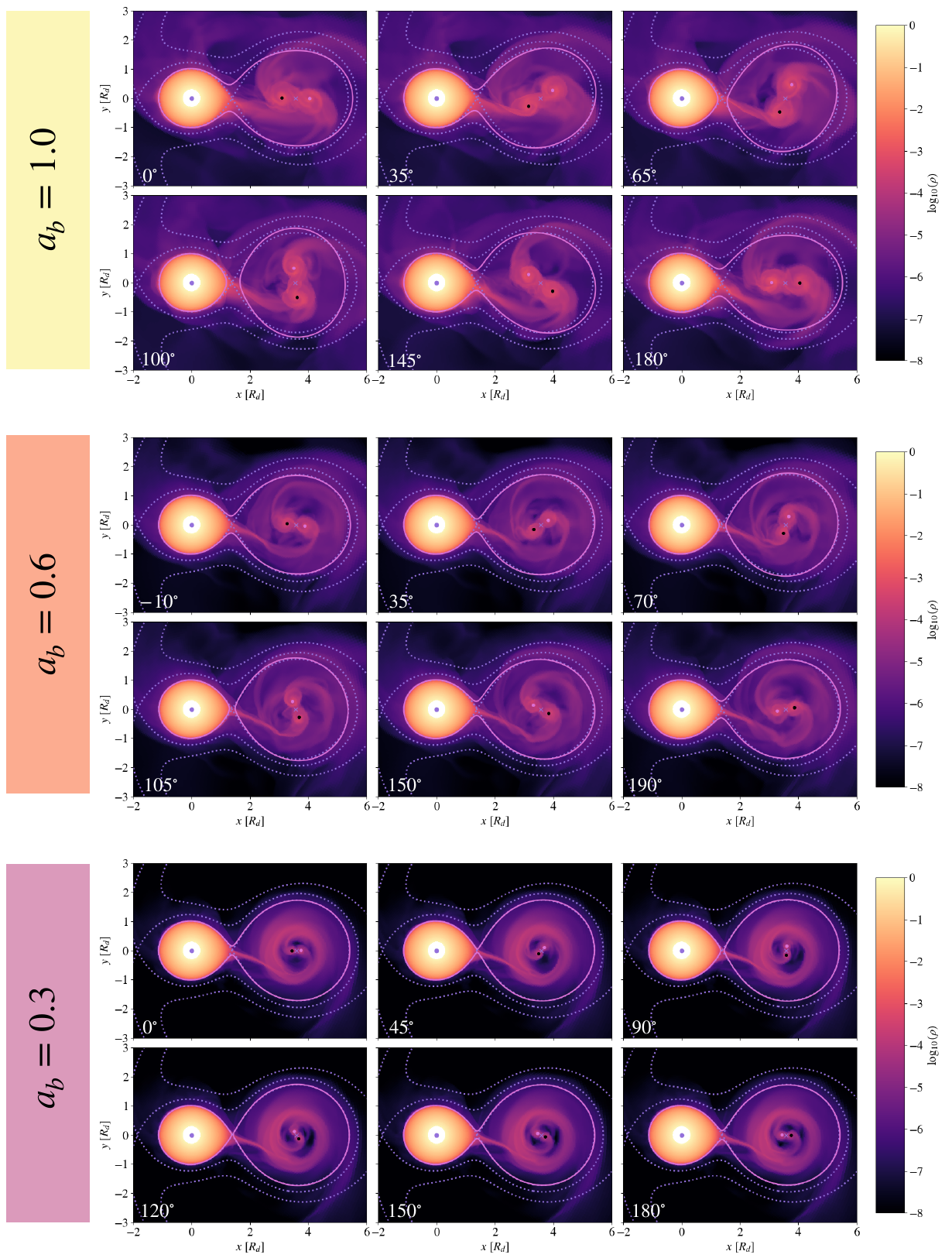}
    \caption{Density snapshots for half of an orbital period of the inner binary, at three different separations $a_b$. Each separation shows 6 different stages of the orbit, starting at the top left and ending at the bottom right of the 6 panels. The Roche equipotentials for the binary case are outlines with the dashed light purple lines and the `x' marks the location of the fictitious single point mass. The light pink continuous lines indicate the calculated equipotantials for the triple case matching the exact phase of the inner binary. The black and pink dots show the two components of the inner binary. The frames are rotated to appear in the corotational frame for ease of comparison. We note that the Roche lobes connect and disconnect as the phase of the inner binary changes throughout its orbit.   \textit{Top:} $a_b = 1.0$. \textit{Middle:} $a_b = 0.6$. \textit{Bottom:} $a_b= 0.3 \ R_d$}
    \label{fig:snapshots}
\end{figure*}

\subsection{Time-dependent Roche geometry}\label{sec:geometry}

Roche lobe overflow episodes are  well studied in the context of binary systems. During these mass transfer episodes, a donor star expands until it reaches the size of its Roche lobe, at which point mass starts to flow though the first Lagrangian point $L_1$ into the Roche lobe of the accretor. In binaries, the Lagrange points are static points in the corotating frame of reference (and, in the inertial frame, these points co-rotate with the system). For triple system though, this does not hold true. The orbital rotation of the inner binary causes the structure of the equipotentials to vary, causing the Roche lobes of the donor and accretor to come in and out of contact. There are no truly static points anymore, static points only occur instantaneously.  Furthermore, the addition of a body to the system creates an additional sixth instantaneous Lagrangian point. 

It is useful to visualize this  system by drawing the outlines for the equipotential surfaces. In Figure \ref{fig:snapshots}, we plot density slices at $z=0$ over the course of half an orbit for three different separations $a_b = 1.0, \ 0.6, \ 0.3$. Each separation is presented in six snapshots from top left to bottom right. We plot these snapshots in the corotating frame to aide comparison. We calculate the equipotential surfaces by assuming a point masses at the location of the donor center and accretor objects. For each  snapshot, we overlay the Roche equipotentials for the binary case, where we take the whole mass of the inner binary $m_b$ to be in a single point mass, with a light purple doted line. We think of this case as the limit where $a_b = 0.0$. The small purple `x' denotes the position of the fictitious point particle $m_b$. We use these Roche lobe equipotentials to calculate a distance $a_t$ at which we will have mass transfer to a single accretor. 

We plot the numerically calculated triple-system equipotential surfaces for the hierarchical triple cases in a light pink, continuous line. We denote each of the components of the accretor binary by a dot, one pink and one black to help illustrate the orbital motion. In the triple case, the Roche lobe of the accretor changes over the course of the inner binary's orbit. We include the measurement of the angle of the inner binary measured from the line that connects the core of the tertiary to the center of mass of the binary. This way $0^\circ$ corresponds to a collinear system where the accretor marked with a black dot is in between the tertiary and the accretor marked by a pink circle. The snapshots are chosen to depict the variability over the course of half an orbit ($180 ^\circ$).

For the largest accretor binary separation in Figure \ref{fig:snapshots}, the Roche lobe goes from highly connected ($0^\circ$, $180^\circ$) to completely detached ($90^\circ$) with changing orbital phase. This pattern remains for the other two cases, albeit in a less extreme fashion. Even for the smallest separation examined in this study, $a_b = 0.15$, even though snapshots are not presented, we note that the qualitative pattern remains. The progression in Figure \ref{fig:snapshots} is consistent with the perturbative picture described in Section \ref{sec:analytical}. At small $a_b$, the donor primarily experiences the monopole component of the inner-binary potential and the Roche geometry approaches the single-accretor limit. As $a_b$ increases, the finite-separation perturbation becomes progressively more important, producing larger excursions of the instantaneous Roche surface.

This behavior provides a direct connection to the perturbative limit discussed in Section \ref{sec:analytical}. When $a_b$ is small, the donor primarily responds to the monopole gravitational field of the inner binary and the Roche geometry approaches that of the single-accretor calculation. Increasing $a_b$ strengthens the leading finite-separation perturbation, producing progressively larger excursions of the instantaneous Roche surface over the inner orbit. The oscillating Roche geometry determines the instantaneous conditions for material to leave the donor, but the resulting mass flux also depends on whether gas near the mass-transfer region can respond to the changing potential.

The overlap between the Roche lobe of the donor and that of the accretor binary widens the transfer nozzle, allowing  mass transfer to occur over a broader stream than when the Roche lobes are in slight in contact. {In Table \ref{table: lengthscales} we present a set of length scales to appreciate the variability of the systems. We calculate the distance from the tertiary to one of the binary components at closest, $a_{t,\rm{near}}$, and farthest, $a_{t,\rm{far}}$ approach. As the inner binary orbits, the Roche lobes will reach greatest overlap at  $a_{t,\rm{near}}$ with a corresponding $L_{1,\rm{overlap}}$ and will be maximally separated after one quarter of an orbit with $L_{1,\rm{separated}}$. When we compare the emerging Lagrangian points to the static $L_{1,t}$ we can see that for larger $a_b$ values, $L_{1,\rm{overlap}}$ is significantly closer and $L_{1,\rm{separated}}$ is significantly farther. Over the course of half an orbit, $L_1$ goes from  $L_{1,\rm{overlap}}$ to $L_{1,\rm{separated}}$ along the axis of $y=0$. } We look at how the overlap of the potentials affect  the flow through $L_1$ in the following section.

\subsection{Hydrodynamic response of the transfer stream}\label{sec:hydro}

\begin{figure*}
    \centering
    \includegraphics[width=0.75\linewidth]{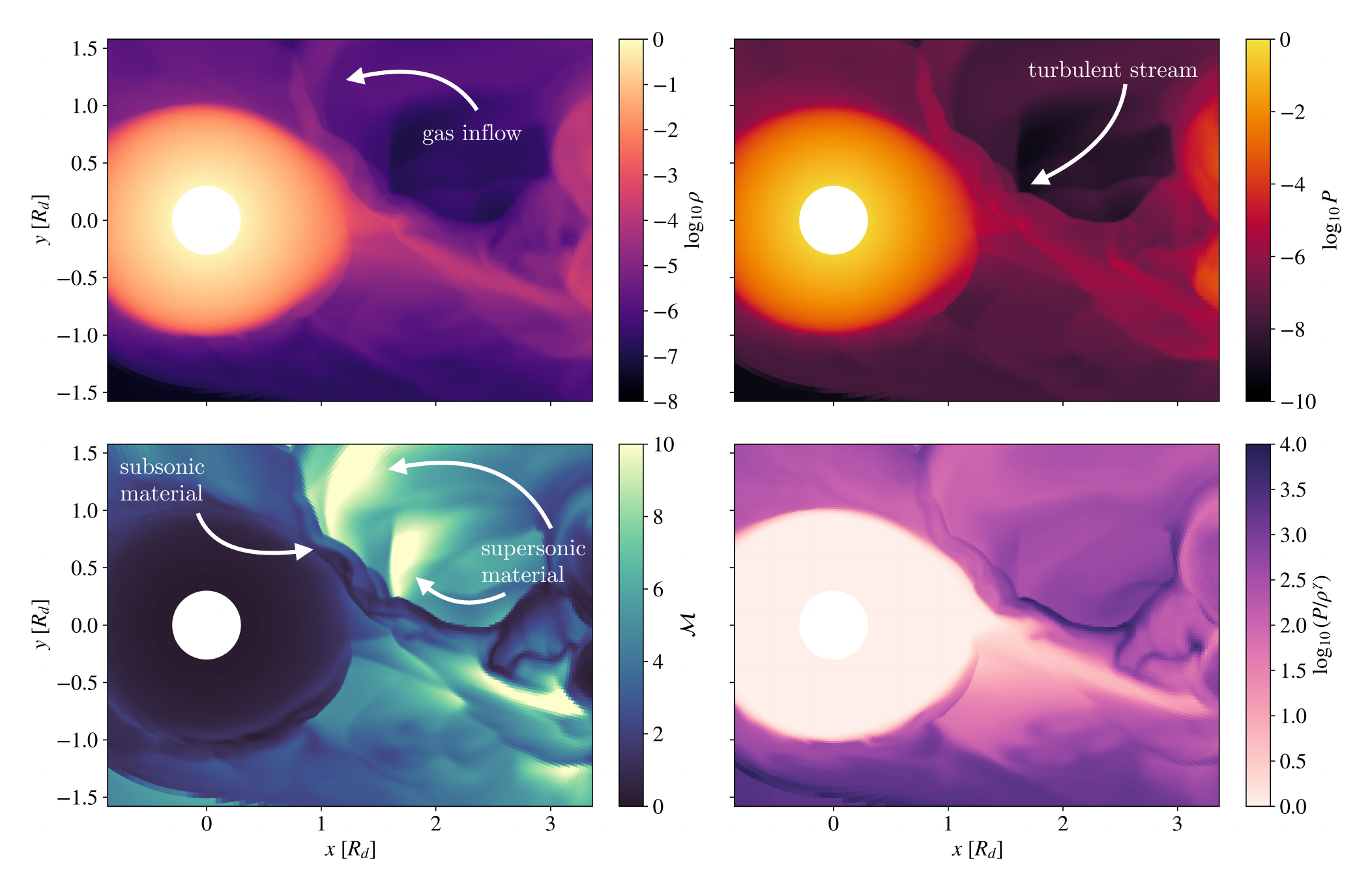}
    \caption{We zoom in around $L_1$ for $a_b = 1.0$ in the corotating frame. At the top left we plot $\log_{10} \rho$, top right $\log_{10} P$, bottom left $\mathcal{M}$, and bottom right $\log_{10}(P/\rho^{\gamma})$. The inner binary (not pictured) is at $90^\circ$ at the time of the snapshot. The mass transfer stream passing through $L_1$ displays a turbulent structure, with gas coming back to the vicinity of the donor after being flung back by the inner binary. There is a clear shock front from supersonic material from the Roche lobe of the inner binary hitting the area of the donor where the mass transfer nozzle lies. The specific entropy highlights the structure where that the gas follows as it passes through $L_1$ towards the inner binary.
    }
    \label{fig:L1_multiplot_1}
\end{figure*}

\begin{figure*}
    \centering
    \includegraphics[width=0.75\linewidth]{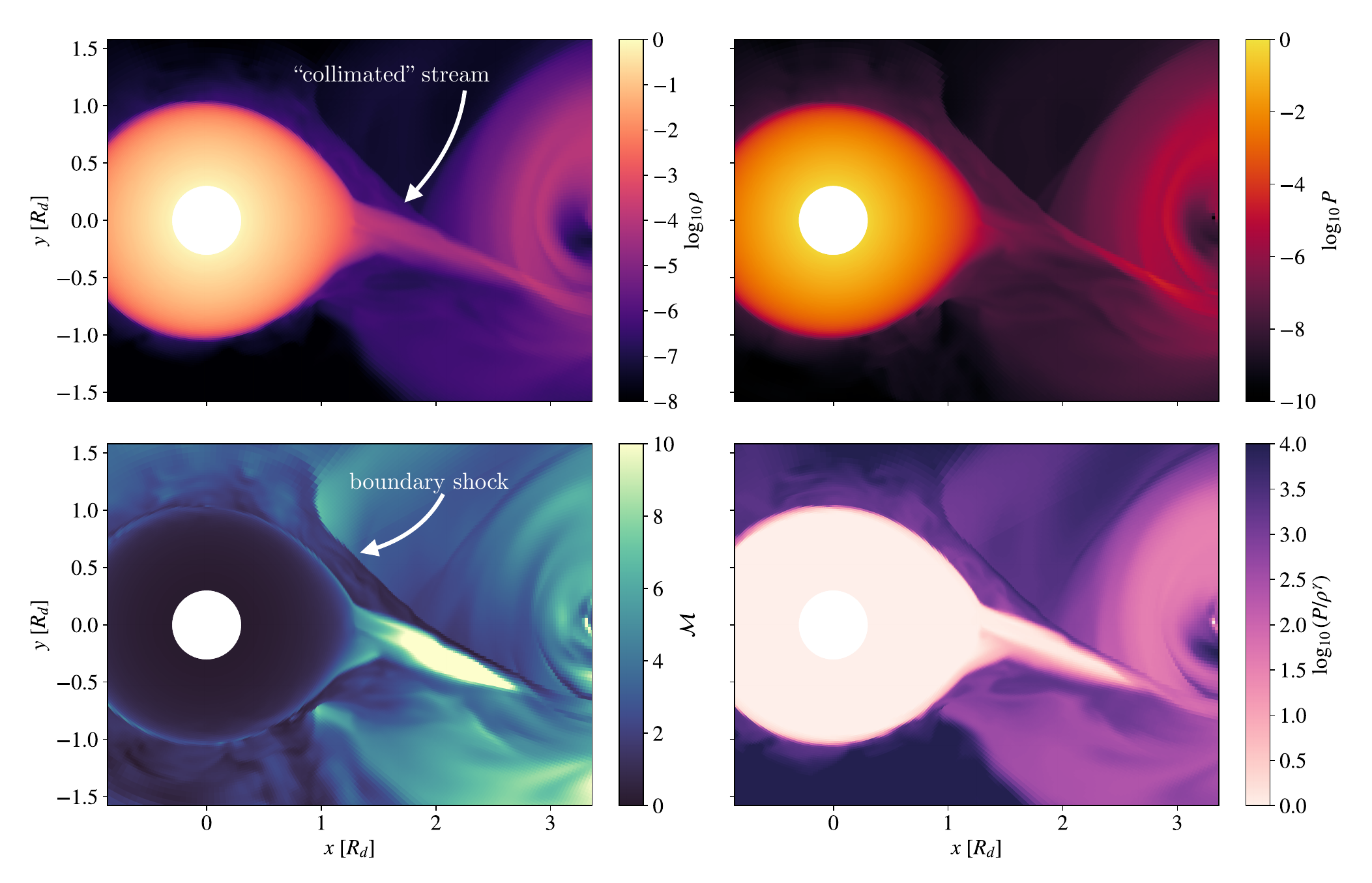}
    \caption{Same setup as Figure \ref{fig:L1_multiplot_1} but now we zoom in  around $L_1$ for $a_b = 0.3$. Here, the mass transfer stream is significantly collimated as it travels through $L_1$. Outside of the Roche lobe structure, there is significantly little material and low pressure. We can still appreciate a boundary shock between the donor and material coming back from the inner binary, but it has little effect on the structure of the tertiary and the mass transfer stream. This allows the material to infall supersonically from $L_1$ towards the inner binary. The specific entropy highlights the structure of the stream and shows that there is little mixing occurring as the material settles onto a circumbinary disk around the inner binary.}
    \label{fig:L1_multiplot_0_3}
\end{figure*}

The donor star fluid is largely corrotating with the outer binary system, and responds to the time-varying potential. Thus, the oscillating Roche geometry determines how and when material can leave the donor. The resulting mass flux also depends on the rate that gas near the transfer nozzle can respond to that forcing. We therefore examine the hydrodynamic structure near $L_1$ for representative wide and compact inner binaries.

In Figure \ref{fig:L1_multiplot_1} we zoom in to focus on the flow near $L_1$ when $a_b = 1.0$ and the inner binary has a $90^\circ$ orientation. At this point in time, the Roche lobes are maximally separated and the inner Lagrangian is located at $L_{1,\rm{separated}}$. In the top left, we look at the density distribution of the material in the vicinity of the stream. We note that there is a widening of the stream compared to the one for a single accretor, accompanied by low density material around the star and the stream. Toward the top of the frame, some of the transferred material is redirected back toward the donor following its interaction with the inner binary. This returning flow illustrates how the  interaction with the inner binary can alter the trajectories of the transferred gas. The orbital acceleration of the inner binary flings material out of its own Roche lobe periodically, with a fraction of this material making it back into the Roche lobe of the tertiary. Another fraction of the material has enough energy to become unbound from the triple system and cool down as it moves radially outwards into the background. This mixing effect can also be appreciated in Figure \ref{fig:snapshots} where one can see that the overall density around the system is significantly higher, at $\rho \sim 10^{-6} - 10^{-7} M_dR_d^{-3}$ than the background density, way under $\rho \sim 10^{-8}M_dR_d^{-3}$. 

In the top right and bottom left panels of Figure \ref{fig:L1_multiplot_1} we plot the pressure and mach number respectively in the region around the mass transfer stream. These plots allow us to better appreciate areas with high pressure differentials and the structure of the shocks. When looking at the pressure (\textit{top right}), we can observe the turbulence on the mass transfer stream caused by the material being flung back towards the tertiary. In the case of $\mathcal{M}$ (\textit{bottom left}), we can see how a region of highly supersonic material (\textit{bright green}) collides with slower material (\textit{dark blue}) at the center top of the panel, driving a shock that moves towards the surface of the donor star and the stream. The bottom right panel plots the specific entropy $P/\rho^{\gamma}$. This allows us to more clearly discern the morphology of the mass transfer stream, as the material flowing from the donor to the binary has relatively low specific entropy while the material around remains at a relative similar specific entropy as it gets mixed around.

We replicate these plots for $a_b = 0.3$ in Figure \ref{fig:L1_multiplot_0_3}. Contrary to the case for $a_b=1.0$, we see that there is limited mixing in density and pressure. The mass transfer stream is more collimated, with a flow that falls onto the bulk material around the binary and remains bound. This creates a much more striking contrast in mach number and entropy, with the highest $\mathcal{M}$ and lowest entropy limited to the stream structure.  We notice a smooth transition through a sonic point around $L_1$ with supersonic infall toward the accretor binary.  

In the wider binary case, some of this structure of the ordinary Roche-lobe overflow mass transfer region remains, but there is also a shocked, high pressure, lower-density and low-mach number boundary layer to the stream that comes from the interaction with the time-varying inner binary potential. In the more compact case, we see a comparatively static boundary shock from the stream-disk interaction - that also does not reach as far back to the donor.

\begin{figure}
    \centering
    \includegraphics[width=0.95\linewidth]{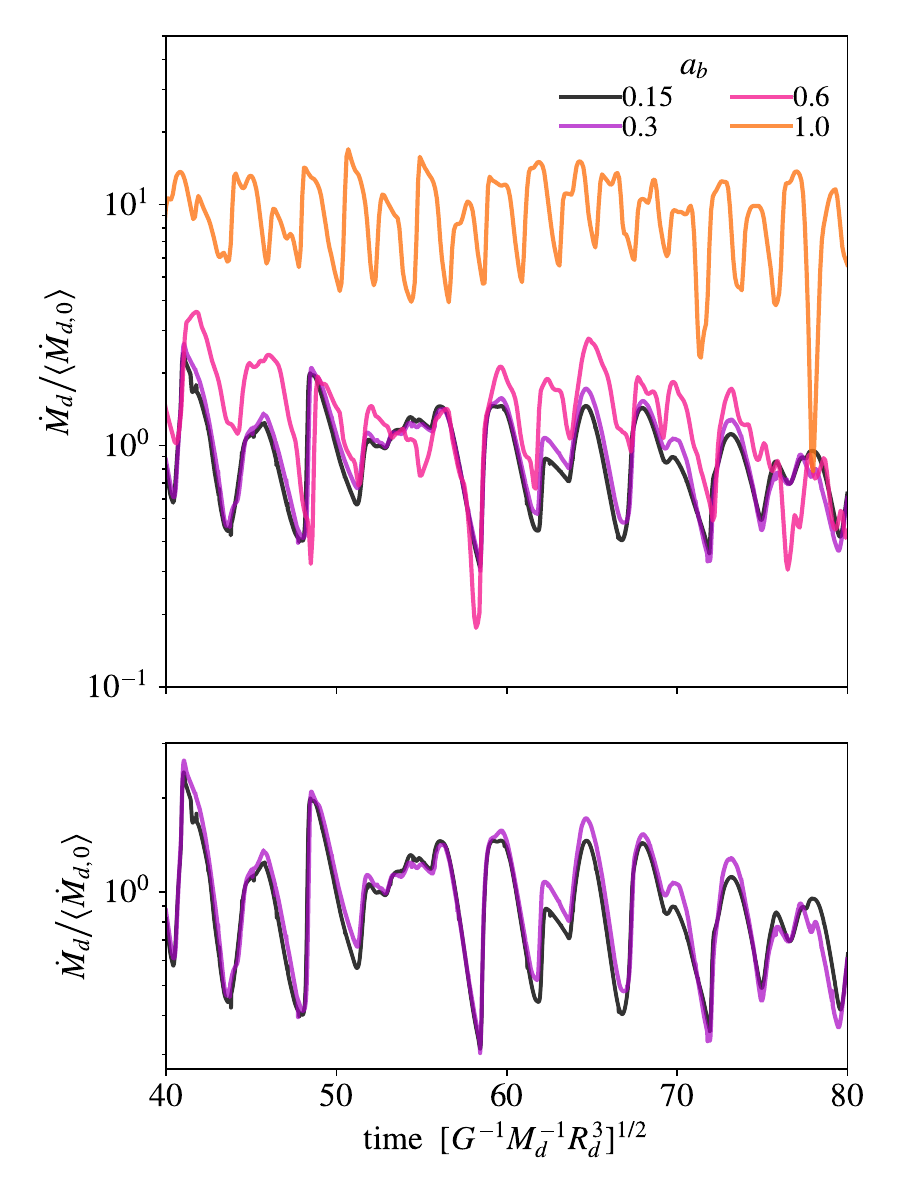}
    \caption{Mass loss rate of the donor star as a function of simulation time. The mass loss rate is calculated as the mass flux out of a spherical surface with a radius $r = 2R_d$ centered about the donor. The negative sign denotes mass loss. \textit{Top:} $\dot{M_d}$ for $a_b = 0.15, \ 0.3, \ 0.6, \ 1.0$ \textit{(black, purple, pink, orange)}.  The mass loss rate has a close-to-periodic variability, with the amplitude of the variablility increasing with the separation of the inner binary. For smaller separations $a_b$, the mass loss rate is close to zero, while for larger separations the mass loss rate is much larger.  \textit{Bottom:} We plot a zoomed in version to appreciate the subtle difference between $a_b = 0.3$ \textit{(purple)} and $a_b = 0.15$ \textit{(black)}. All values are normalized by the average mass loss rate of a tertiary with a single accretor. }
    \label{fig:mdot_time}
\end{figure}

\begin{figure}[]
    \centering
    \includegraphics[width=0.95\linewidth]{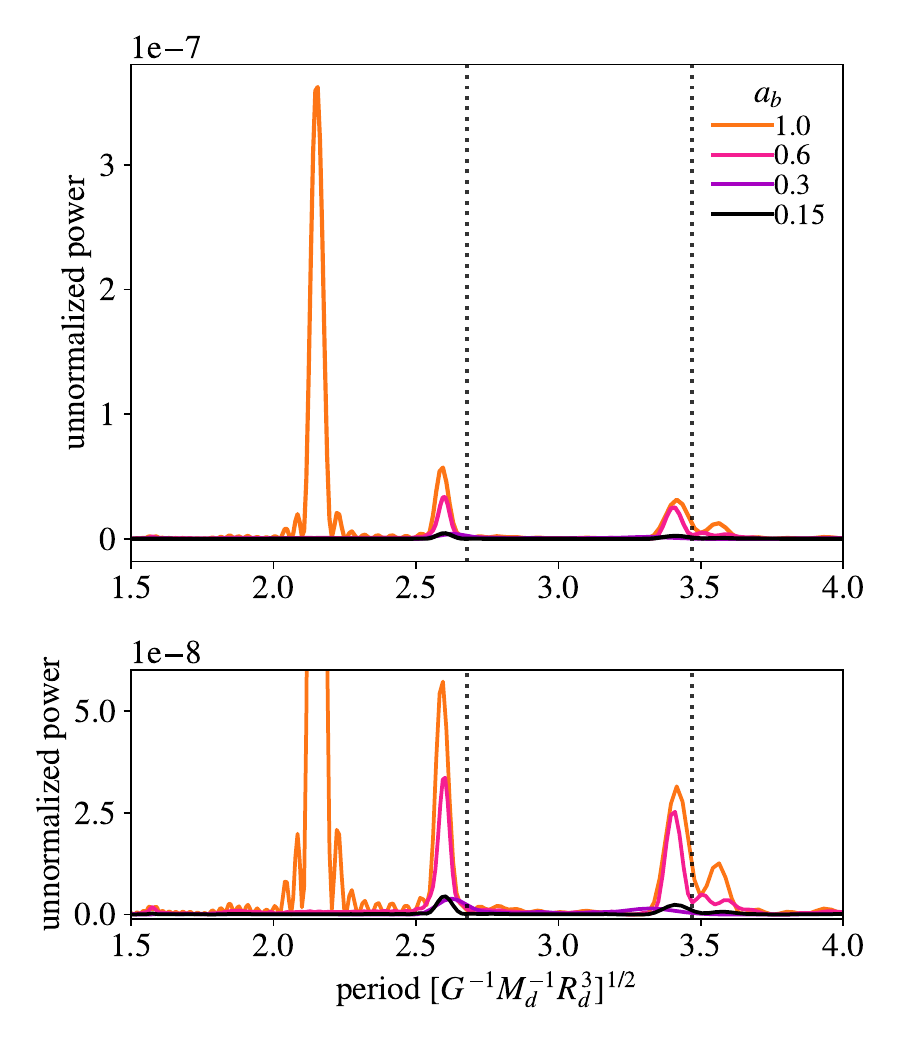}
    \caption{Lomb-Scargle Periodogram of the mass flux at a spherical surface with $r=1.1 R_d$. We plot, from darker to lighter, the period decomposition with ascending $a_b$. The vertical dotted lines demark the characteristics period predictions at $T=2.7$ and $T=3.47$ in units of $(G^{-1} M_d^{-1} R_d^3 )^{1/2}$.}
    \label{fig:periodogram}
\end{figure}

\subsection{Time-dependent mass transfer \& forced donor oscillations}\label{sec:oscillations}

In the traditional binary Roche lobe overflow prescription, mass that flows into the Roche lobe of the accretor does not have sufficient energy to get out of it nor leave the gravitational influence of the system. This leaves the gas with two options; collide with the accretor or settle into a disk around it. The modification of having a binary accretor changes this assertion. The orbit of the inner binary transfers angular momentum to the gas, exciting the material and allowing it to escape. The widening of the stream enables material from innermore layers of the star to flow into the Roche lobe of the binary accretor. The combination of these two effects has profound implications on the mass loss rate of the donor tertiary. 

We plot the rate of mass loss of the donor, $\dot{M}_d$, as a function of time and inner binary separation in Figure \ref{fig:mdot_time}. We calculate the mass loss as the flux of material out of a spherical surface with radius $2R_d$  from the center of the donor. Negative $\dot{M}_d$ values denote material flowing away from the tertiary. We normalize $\dot{M}_d$ to the average mass loss rate of a single, $\langle \dot{M}_{d,0} \rangle$, which is $\sim -8.2 \cdot 10^{-7}$ in code units of $(GM_d^3R_d^{-3})^{1/2}$. In the case of wider inner binaries, we can appreciate a quasi-periodic behavior, with larger average $\dot{M}_d$ values and larger amplitude in variability. For more compact inner binaries, the donor loses little to no mass over time. In the bottom panel, we zoom in around $\dot{M}_{d}/\langle \dot{M}_{d,0} \rangle=1$ in order to visualize the subtle difference in the mass loss rate for $a_b = 0.15$ and $a_b=0.3$. There are still variations in time for both binary separations, but they seem to stabilize around $\dot{M}_{d}/\langle \dot{M}_{d,0} \rangle=1$, very close to the average value for a single accretor.

The strong time variability in Figure \ref{fig:mdot_time} also raises a separate question: does the donor respond only through the changing mass-transfer geometry, or does the periodic forcing of the inner binary couple to the donor’s intrinsic oscillations? With two kinds of orbits in a hierarchical triple, we expect some natural frequencies to arise in the system. We investigate these frequencies by performing a Lomb-Scargle periodogram on the mass flux distribution over time. The Lomb-Scargle periodogram reveals the strongest periods that arise in each one of the simulations in our study.  We calculate this flux at a distance $1.1 R_d$, which highlights the frequencies of the mass transfer from the donor over time. The different binary separations $a_b$ are each plotted in a different shade of blue. We present the period decomposition in Figure \ref{fig:periodogram}.

In the top panel we notice a very prominent peak at a period of around $2.15$ in code units of $(G^{-1} M_d^{-1} R_d^3 )^{1/2}$ for the case of $a_b = 1.0 R_d$. This peak is not replicated for any of the other inner binary separations. The bottom panel shows a zoomed in version where the rest of the peaks are more easily appreciable. Here, we can make out two more peaks that arise, to some level, in all of the binary separations. The height of these persistent peaks increases with binary separation, suggesting a stronger excitation of the donor's intrinsic oscillations. We identify two prominent periods at $T\approx2.6$ and $T\approx3.4$, in units of $(G^{-1}M_d^{-1}R_d^3)^{1/2}$. 
These are consistent with the characteristic periods $T=2.7$ and $T=3.47$ predicted by \texttt{GYRE} \citep{gyre} for the radial $(\ell=0,m=0)$ and axisymmetric quadrupolar $(\ell=2,m=0)$ modes of our donor model, respectively.
These periods are plotted as vertical dotted lines in both panels of Figure \ref{fig:periodogram} for comparison.
The periods measured in the simulations are slightly shorter than those predicted by \texttt{GYRE} for the spherical donor model, differing by approximately $2$-$4\%$. These small differences may reflect changes in the donor's equilibrium structure when mapped onto the numerical grid, as well as rotational and tidal deformation not fully captured by the spherical stellar model. We do not attempt to isolate these contributions here, but the close agreement supports identifying the observed peaks with the donor's intrinsic oscillation modes.


Coming back to the first peak around $T \sim 2.15$ $(G^{-1} M_d^{-1} R_d^3 )^{1/2}$, we attribute this frequency to half of the rotational orbital period of the inner binary in the inertial reference frame centered at the middle of the tertiary. We note that this period is different from the half-period of the inner binary in the corotating frame. We do not see a comparable peak for the half-orbital period of any of the other separations. This is likely due to the range of the effect that the inner binary has on the gas and the tertiary. For $a_b = 1.0 R_d$ we see that the inner binary forces material back into the Roche Lobe of the tertiary. This is clearly displayed in Figure \ref{fig:L1_multiplot_1} and discussed in the previous sections. In the pressure plot (top right) we can see a higher pressure front coming in from the top right onto the tertiary. This same pattern is shown in the $\mathcal{M}$ plot (bottom left) where a supersonic front is pushing subsonic material, almost squeezing it, onto the tertiary. In contrast, smaller separations do not have the ability to bring back material onto its Roche Lobe of origin.

{We compare the reaction timescale of the mass transfer stream, $\tau_{\rm{reac}}$ to half the period of the inner binary, $T_b$, to see whether the steam has the time to react throughout the orbit. The morphology of the stream changes dramatically with $a_b$ but its width, $w_s$, remains fairly constant at around $0.5R_d$. By defining the reaction timescale as}

\begin{equation}
    \tau_{\rm{reac}} = w_s/c_s
\end{equation}

\noindent{where we are approximating the reaction velocity as the sound speed $c_s$ of the gas. This timescale is essentially the sound crossing time of the stream. The value of $c_s$ in the donor depends on the radial distance from the core. At the depth where the stream opens up, near the surface of the star, $c_s\approx0.4 \ (GM_dR_d^{-1})^{1/2}$. This gives us a roughly approximated value of $\tau_{\rm{reac}} \approx 1.25$. We compare this value to the half-orbital period of the inner binary in the rotating frame of reference, $T_b = [0.11, 0.31, 0.91, 2.18]$ in code units of $(G^{-1} M_d^{-1} R_d^3 )^{1/2}$ for $a_b = [0.15, 0.3, 0.6, 1.0] R_d$. For all but $a_b = 1.0 R_d$, $T_b < \tau_{\rm{reac}}$, meaning that the the change in the potential generated by the orbit of the inner binary on the stream happens faster than the stream can react. Essentially, the stream is minimally affected by the variations in the potential induced by the inner binary. Only for $a_b = 1.0$ we have the case where $\tau_{\rm{reac}}$ is smaller, so this is the only case where the stream reacts to the changing potential and its morphology varies over the course of an orbit.}

\begin{figure}
    \centering
    \includegraphics[width=0.9\linewidth]{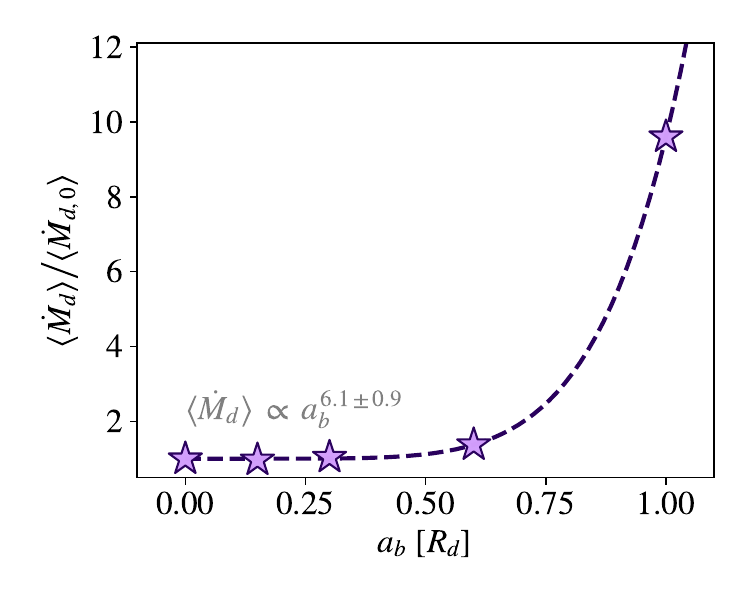}
    \caption{Mass loss rate of the donor star normalized to the mass loss rate for the case of a single accretor $(a_b = 0.0)$. Mass loss rate is calculated as the mass flux out of a spherical surface with a radius $r = 2R_d$. For smaller separations $a_b$, the mass loss rate is the same as the case of a single accretor. For larger separations, the mass loss rate increases dramatically as $a_b^{6.1}$. The fit is done as $\langle \dot{M}_d \rangle = \langle \dot{M}_{d,0} \rangle \left(1+ C \ a_b^{\alpha} \right)$, with $C$ as a proportionality constant, to model the behavior at all studied separations. All values are normalized by the average mass loss rate of a tertiary with a single accretor.} 
    \label{fig:mdot}
\end{figure}

\subsection{Time-averaged mass transfer}\label{sec:m_dot}

To further understand the behavior of the mass loss of the donor, we calculate the orbit averaged $\dot{M}_d$ for the different inner binary separations. In Figure \ref{fig:mdot} we plot our findings for $\langle \dot{M}_d \rangle$ as a function of the inner binary separation. As expected, for small values of $a_b$, the mass loss rate behaves as that of a single accretor. This is not surprising as the structure of the stream and the disk around the binary accretors closely resembles those of the single accretor. As the separation of the inner binary increases, $\langle \dot{M}_d \rangle$ increases substantially as $a_b^{6.1\pm0.9}$ when we fit the orbit-averaged mass loss rate as $\langle \dot{M}_d \rangle = \langle \dot{M}_{d,0} \rangle \left(1+ C \ a_b^{\alpha} \right)$, with $C$ as a proportionality constant. The finite separation of the inner binary affects the donor mass-loss rate through the changing Roche geometry. At larger $a_b$, the overlap of the Roche lobes also permits mass loss through the outer Lagrangian region, so that the donor can lose material from opposite sides. After material enters the inner-binary Roche volume, the orbital motion of the binary further redistributes its energy and angular momentum.


The steep dependence of the mass-loss rate on $a_b$ can be understood qualitatively from the perturbative structure of the binary potential. For the equal-mass binary considered here, the leading correction to the point-mass potential is quadrupolar, such that the
perturbation near the mass-transfer region scales as

\begin{equation}
    \delta \Phi \propto a_b^2.
\end{equation}

\noindent Roche-lobe overflow responds nonlinearly to the effective potential excess above the critical Roche surface. For a polytropic atmosphere, the mass flux scales approximately as

\begin{equation}
    \dot{M} \propto (\Delta \Phi)^{n+3/2}
\end{equation}

\noindent \citep{1972AcA....22...73P}. With the polytropic index $n=3/2$ donor adopted here, this yields

\begin{equation}
    \dot{M} \propto (\Delta \Phi)^{3}.
\end{equation}

\noindent If the additional effective overfill is controlled primarily by the finite-binary perturbation, then

\begin{equation}
    \Delta \Phi_{\rm{pert}} \propto a_b^2,
\end{equation}

\noindent and the enhanced component of the mass-loss scales as

\begin{equation}
    \dot{M}_{\rm{pert}} \propto \left( a_b^2 \right)^3 = a_b^6.
\end{equation}

\noindent This scaling is consistent with the numerically measured dependence in Figure \ref{fig:mdot}. 

For an arbitrary polytropic index $n$, combining the quadrupolar scaling of the finite-binary perturbation with the polytropic Roche-lobe overflow response gives

\begin{equation}
    \dot{M}_{\rm pert}\propto a_b^{\,2n+3}. 
\end{equation}
The exponent is therefore not universal but depends on the donor’s structure and mass-radius response through the value of  $n$. Thus donors with different structures and mass-radius responses will exhibit different dependence on $a_b$.

This scaling applies to the enhanced component of the mass-transfer rate when the finite-binary perturbation controls the additional Roche-lobe overfill. For the general case of $ \Delta\Phi(t)=\Delta\Phi_0+\Delta\Phi_{\rm{pert}}(a_b,t),$
 the \(a_b^{2n+3}\) dependence should not be expected to extend into the compact-binary limit. Instead, as seen in Figure \ref{fig:mdot}, the mass-loss rate approaches the single-accretor value as \(a_b\rightarrow0\).

We do not expect this scaling to remain exact across arbitrary binary separations or donor structures. Rather, it provides a perturbative interpretation of why a relatively modest change in the binary potential can produce a much steeper response in the donor mass-loss rate. The enhanced donor mass loss therefore appears to originate from the nonlinear response of Roche-lobe overflow to the time-dependent binary potential. After material crosses the $L_1$ point, gravitational torques from the inner binary further redistribute its energy and angular momentum, determining whether material remains bound, returns toward the donor, or escapes the system.






\subsection{Individual versus circumbinary accretion}\label{sec:ind_vs_cbd}

Having established how inner-binary separation regulates the supply of material from the donor, we now consider the fate of that material after it enters the Roche lobe of the inner binary. The morphology of the gas around the inner binary shows a clear transition between each component of the inner binary accreting individually $(a_b = 1.0, 0.6)$ and them accreting collaboratively onto a circumbinary disk $(a_b = 0.3, 0.15)$. We attribute this to two competing behaviors, the angular momentum conservation from the material at rest at $L_1$ which circularizes at $a_{\rm{circ}}$ and the angular momentum deposition from the inner binary which tries to open a cavity at $a_{\rm{cav}}$. In Table \ref{table: lengthscales} we presented the theoretical values for these two length scales.

In order to sustain a circumbinary disk, $a_{\rm{circ}}$ must be larger than $a_{\rm{cav}}$, and, by definition, also larger that $a_b$ itself.
For our whole suite of simulations, $a_{\rm{circ}} = 0.65$. This value draws a transition between individual and collaborative accretion at  $a_{b} \approx 0.39$. The results of our simulations align with these approximation, as we begin to observe slight collaborative accretion for the $a_b = 0.3$ case.  Figure \ref{fig:snapshots} shows this transition. For the top two cases ($a_b = 1.0$ and $a_b=0.6$) there is a defined mass transfer stream that directly connects the $L_1$ point to the bottom binary component for the counterclockwise inner binary orbit (i.e. the component with a negative $y$ coordinate moving away from the tertiary). After half an orbit, when the tertiary and binary elements are collinearly aligned at $y=0$, the individual accretion stream transitions from one element to the other. 

At smaller $a_b=0.3 R_d$ and below, we instead see $a_{\rm{cav}}$ and $a_b$ fall inside of $a_{\rm{circ}}$. The mass transfer stream connects to a single circumbinary structure contained within the inner binary's Roche lobe. The angular momentum of the binary is able to clear some of the material within the circumbinary structure, forming a cavity inside of the disk. For this specific case, we also see the emergence of disk-like structures surrounding each of the individual components. These `mini disks' show that the binary is not sufficiently efficient to completely clear the material within the circumbinary disk cavity. This behavior is heavily studied in the context of protoplanetary disks \citep{M_sta_2019, 2019ApJ...871...84M} and circumbinary disks around binary black holes \citep{2014ApJ...783..134F, Thun_2017, 2019ApJ...884...22A,2020ApJ...889..114M, 2020ApJ...900...43T,2023ApJ...944...44K,2024ApJ...977...16R,2025ApJ...982L..11H,2026ApJ...997...13R,2026ApJ..1004..196D} for both thick and thin disks. The long term evolution of the inner binary and the disk has been shown to depend in the orbital parameters, mass and eccentricity, of the binary opening the cavity \citep{2024A&A...688A.128V}. Material that remains within the cavity will settle into a smaller orbit around one of the binary elements. Interestingly, we do not see this behavior arise for $a_b = 0.15$ (not pictured in the snapshots) because at this separation, it is hard to distinguish the inner boundary of the disk due to our resolution and the fact that the gas is very near the softening radius of the point masses.   

\begin{figure*}
    \centering
    \includegraphics[width=0.95\linewidth]{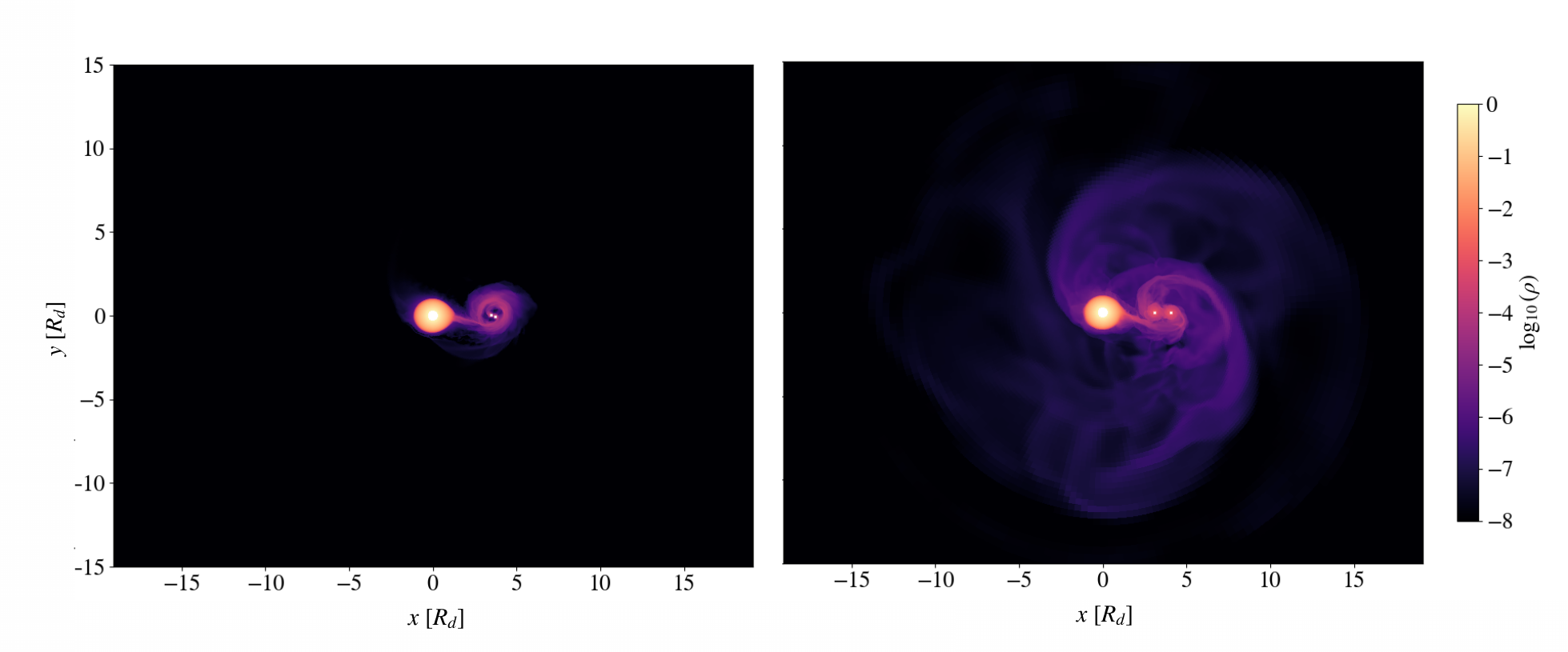}
    \caption{Zoomed out snapshots of the density distribution at late time for triples with $a_b = 0.3$ (left) and $a_b = 1.0$ (right). For $a_b = 1.0$ we can observe the amount of material being flung out by the inner binary and by the tertiary through its outer Lagrangian point. The material takes the shape of spiral arms given that the simulations are performed in the inertial frame of the tertiary. These snapshots are rotated to appear in the corotational frame to facilitate comparisons and visualize the overall structure.}
    \label{fig:zoom_out}
\end{figure*}

\subsection{Outflows and global morphology}\label{sec:global}


In this subsection we take a look at the global morphology of the gas surrounding the triple system. We present to snapshots at $y=0$ for $a_b = 0.3$ (left) and $a_b = 1.0$ (right) in Figure \ref{fig:zoom_out}. 

From the previous analysis of the distribution of density about the Roche lobes, we can anticipate whether gas would be ejected from the system or remain bound around the inner binary. For short binary separations, we note that the material settles into a disk-like structure within the Roche lobe of the accreting binary. The angular momentum of the binary is sufficient to clear a cavity inside this disk, but not sufficient to transport material out of the Roche lobe. When we look at the structure of this case from afar, we see a pattern similar to the case for a single accretor Roche lobe overflow. There is a stream of material that flows from the donor onto a disk formed around the accretor. The majority of the material is contained within the Roche lobes.

The opposite behavior takes place for large $a_b$ values. In this case, the binary separation is too large to sustain a disk, and instead, the binary pushes material out of its own Roche lobe. Some of this material comes back to the donor's Roche lobe, but the rest is ejected away. The outflowing material forms spiral structures as the inner binary  rotates with the outer orbit. These structures contain contributions from both material redistributed and ejected by the inner binary and material lost through the donor's outer Lagrangian point. The donor's outer Lagrangian outflow is caused by the oscillations of the potential and the degree of overlap between the Roche lobes.  This results in a pulsed mass ejection from the binary's side (with the pulses corresponding to half periods of the inner binary) and a continuous mass ejection from the outer Lagrangian on the tertiary's side.

The local differences in mass-transfer rate and accretion morphology produce qualitatively different outcomes on the scale of the entire triple. Compact inner binaries retain much of the transferred material in a circumbinary structure, whereas wider binaries more efficiently redistribute and eject the gas supplied by the donor.

%


%
%





\section{Summary and Conclusions}\label{sec: conclusions}

In this study, we explore the onset of tertiary mass transfer by performing a suite of hydrodynamic simulations with varying inner binary separation $a_b$. We focus on systems with constant mass ratio $q=M_d/m_b=1/3$ and begin the simulations at the point where Roche lobe overflow commences. We look at the effect of inner binary separation on the mass transfer rate, stellar oscillations and possible circumbinary disk formation in hierarchical triple systems.  The summary of our findings are below:

\begin{itemize}
    \item For large values of $a_b$, the Roche lobes of the donor and inner binary overlap significantly, allowing material to be exchanged between them and, in some cases, ejected from the system. The escaping material forms extended spiral structures (Sections \ref{sec:geometry}, \ref{sec:hydro} \& \ref{sec:global}; Figures \ref{fig:snapshots}, \ref{fig:L1_multiplot_1}, \ref{fig:L1_multiplot_0_3} \& \ref{fig:zoom_out}).
    \item The donor retains signatures of its intrinsic oscillation spectrum, while the widest inner binary introduces an additional periodic response associated with binary forcing. The appearance of this response suggests that the relationship between the forcing period and the donor's hydrodynamic response time plays an important role in determining how the donor responds to the changing potential (Section \ref{sec:oscillations}; Figures \ref{fig:mdot_time} \& \ref{fig:periodogram} ).
    \item The orbit-averaged donor mass-loss rate approaches the single-accretor result for compact inner binaries but increases steeply with binary separation, approximately as $\langle\dot M_d\rangle\propto \langle\dot M_{d,0}\rangle\left(1+C\  a_b^{6.1\pm0.9}\right)$ over the range explored in our simulations. This behavior is consistent with the nonlinear response of Roche-lobe overflow to the leading quadrupolar perturbation of the inner-binary potential (Section \ref{sec:m_dot}; Figure \ref{fig:mdot}). For a polytropic donor, the perturbative argument gives $\dot M_{\rm pert}\propto a_b^{2n+3}$, so the approximately sixth-power dependence found here is specific to our $n=3/2$ donor rather than universal.
    \item We identify a transition from individual accretion onto the inner-binary components to collective accretion through a circumbinary structure near $a_b\approx0.4$. This transition reflects the competition between gas circularization and gravitational torques from the inner binary, which can clear a central cavity (Section \ref{sec:ind_vs_cbd}; Figure \ref{fig:snapshots}).

\end{itemize}

Taken together, these results show that inner-binary separation heavily impacts tertiary mass transfer. In the compact limit, the binary approaches a point-mass accretor, whereas increasing its separation modifies the Roche geometry, amplifying the donor response, and changing whether the transferred material is retained, redistributed, or ejected. The magnitude of the mass-transfer response depends not only on the binary geometry but also on the structure and mass-radius response of the donor.

\begin{acknowledgements}
We are grateful for conversations and exploration with Rosanne Di Stefano and Sophie L. Schr{\o}der that launched this work. We additionally thank Xiaoshan Huang, Sierra Dodd, Nick Kaaz, and Melinda Soares-Furtado for helpful discussions. The UCSC team is supported in part by the Heising-Simons Foundation, the Vera Rubin Presidential Chair at UCSC, and the National Science Foundation (AST-2447606, AST-2307710, AST-2206243, AST-2150255 and AST-1911206). This research was also made possible through funding from the Lamat Institute \citep{2025NatAs...9.1770Q}.
We acknowledge use of the lux supercomputer at UC Santa Cruz, funded by NSF MRI grant AST 1828315. M.M. acknowledges the use of Claude Sonnet 5.5 to compare the Athena++ source code to manuscript equations to ensure accuracy of reported methods section. E.R.-R. used OpenAI Codex \citeyearpar{OpenAI2025_Codex}\! for assistance with manuscript editing.
\end{acknowledgements}

\software{ Athena++ \citep{2020ApJS..249....4S}, Gyre \citep{gyre}}


\begin{appendix} 

\section{Circumbinary Disk Harmonics}\label{app: harmonics}

In Section \ref{sec:cavity} we present the values of the cavity induced by the inner binary as it orbits. Here we expand the calculations of the harmonics relative to our problem, closely following Section 2 of \cite{cavity_radius_1994}. 

The Fourier decomposition of the time-periodic gravitational potential of the binary, ${\Phi}$ gives a perturbative potential, 

\begin{equation}
    {\Phi} (r, \varphi, t)
    =
    \sum_{m\ell} \phi_{m\ell}(r) \ 
    \text{exp}\left[ i(m\varphi - \ell {\Omega}_{\rm{in}} t)\right],
\end{equation}

\noindent with azimuthal harmonic number $m\geq0$, time-harmonic number $\ell$ , and  radially variable potential component $\phi_{ml}(r)$. This is calculated using mean motion of the binary ${\Omega}_{\rm{in}} = (Gm_b/a_b^3)^{1/2}$. Each combination of $(m,\ell)$ gives rise to a set of Lindblad resonances, with $\phi_{m\ell}$ giving the respective value for the potential.

We can express $\phi_{m\ell}$ for each component $j$ of the binary as, 

\begin{equation}
    \phi^{(j)}_{m,\ell} (r) 
    =
    (-1)^{mj+1} \frac{Gm_j}{r} 
    \int_0^\pi \frac{d({\Omega}_{\rm{in}}t)}{\pi} 
    \int_0^\pi \frac{d\varphi}{\pi} T(t)B(\varphi,t).
\end{equation}

\noindent Here, $j$ takes the value of $1 \ (2)$ for the binary components $m_1$ ($m_2$). $T(t)$ and $B(\varphi, t)$ are defined as

\begin{equation}
    T(t) = \cos [(m-l){\Omega_{\rm{in}}}t \ + \ m(f-{\Omega}_{\rm{in}}t)], 
\end{equation}

\noindent and

\begin{equation}
    B(\varphi, t) = \frac{\cos m\varphi}{ [1-2q_j(t) \cos \varphi + q_j(t)^2]^{1/2}. }
\end{equation}

\noindent with $\varphi_j=\varphi- f- \pi(2-j)$ being the angle between \textbf{r} and $\textbf{r}_j$, where $\textbf{r} = (r, \varphi)$ and $\textbf{r}_j = (r_j, \varphi_j)$ denotes the position of the $j$th component, and $q_j(t) = r_j(t)/r$ being the radial ratio. Looking at equations 42, 43, 47, 50 in the Appendix of \cite{cavity_radius_1994}, in which they integrate out the time dependence, we get the following relations between the potentials $\phi$ and the eccentricity of the orbit $e$, 

\begin{equation*}
    \phi_{m,m}^{(j)} \propto \frac{1}{r} + \frac{e^2}{r}, 
\end{equation*}

\begin{equation*}
    \phi_{m,m\pm 1}^{(j)} \propto \pm \frac{e}{r} + \frac{e^3}{r}, 
\end{equation*}

\begin{equation*}
    \phi_{m,m\pm 2}^{(j)} \propto \frac{e^2}{r}, 
\end{equation*}

\noindent and

\begin{equation*}
    \phi_{m,m\pm 3}^{(j)} \propto \frac{e^3}{r}. 
\end{equation*}

\noindent These allows us to conclude that only terms with $m=\ell$ remain for a circular orbit with $e=0$. This yields the following potential,

\begin{equation}
    \phi_{m,m}^{(j)} = (-1)^{mj+1} \frac{Gm_b}{2 r} 
    b_m(q)
\end{equation}

\noindent where

\begin{equation}
    b_m(q) = 
    \frac{2}{\pi } 
    \int_0^\pi \frac{\cos \ m \varphi}{(1-2q_j\cos \ \varphi +q_j^2)^{1/2}}d\varphi
\end{equation}

\noindent Summing over the two binary components we have. 

\begin{equation}
    \sum_j \phi_{m,m} = \frac{Gm_b}{2 r} 
    b_m(q) \left[ (-1)^{m+1}   + (-1)^{2m+1} \right]. 
\end{equation}

\noindent The second term in the square brackets always has an odd exponent, therefore it is always $-1$. For $\sum_j \phi_{m,m}$ to be nonzero, we require the first term in the square brackets to have an odd exponent, making $m$ even. This means that the $(m,\ell)=(1,1)$ harmonic does not exist. The next leading nonzero harmonic is then $(m,\ell)=(2,2)$. Now that we know which is the leading harmonic for our setup, we can calculate the radii of the Lindblad resonances using Equation 10 from \cite{cavity_radius_1994}, 

\begin{equation}
    r_{_{LR}} = [(m+1)/\ell]^{2/3}a_b.
\end{equation}

\noindent Solving for $(m,\ell)=(2,2)$ we get that the leading resonance occurs at $1.31 \ a_b$.

\cite{cavity_radius_1994} carry out smooth-particle hydrodynamic simulations that account for the viscosity of the disk. These results are close to the analytically calculated value above, but they find that the cavity opened has a radius of $1.68 \ a_b$. For the study, we use the numerical value of the cavity radius opened by the inner binary.

\begin{figure}[t]
    \centering
    \includegraphics[width=0.95\linewidth]{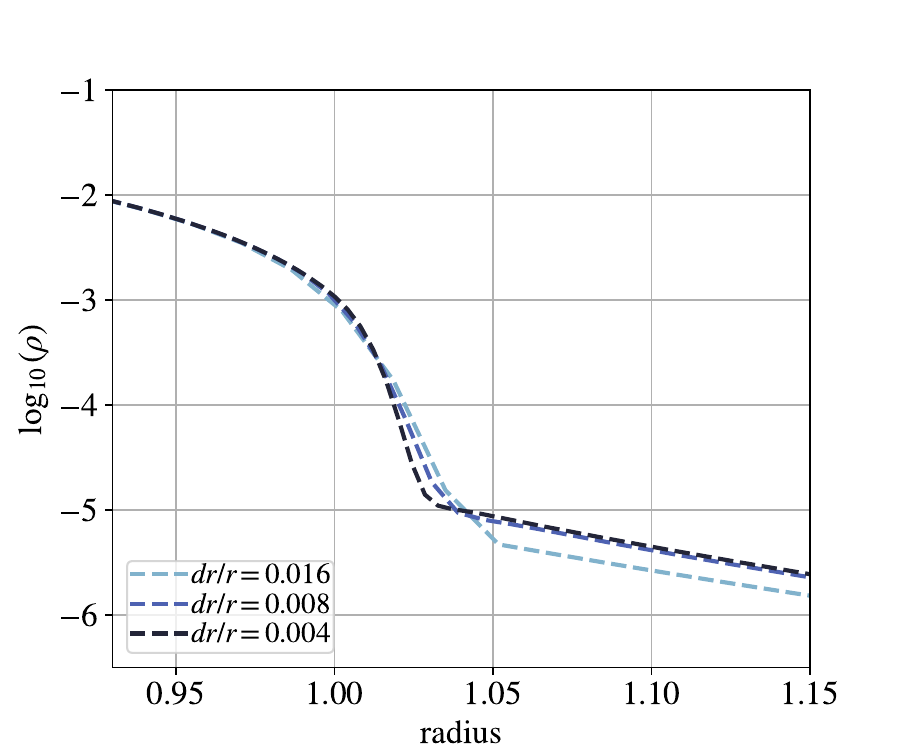}
    \caption{One-dimensional tests of hydrostatic equilibrium at varying spatial resolution while keeping $dr/r$ constant in units of $R_d$.}
    \label{fig:1d_convergence}
\end{figure}

\section{Convergence Testing}\label{app:convergence}

The most numerically stringent component of our simulations is the ability to maintain the structure of the giant star in hydrostatic equilibrium over the duration of the simulations. This is essential because we want to monitor the response to the external forcing of the accretor objects, not loss of intrinsic hydrostatic structure. The density and pressure scale heights towards the edge of the star are very small, requiring a high resolution to resolve them properly. Low resolution approaches can potentially disturb the structure of the star, extending the atmosphere or disbanding the envelope altogether. 

Keeping this in mind, we chose a spherical polar coordinate system centered on the donor star in our simulation. We conducted a set of convergence tests in 1D, 2D, and 3D in order to asses the stability of the donor star at different resolutions. In Figure A\ref{fig:1d_convergence} we present the 1D  tests of hydrostatic equilibrium of an envelope with no perturber objects. The profile shown is taken along the midplane where $\theta=\pi/2$ after $100$ dynamical timesteps and the zones are logarithmically scaled in the radial direction. This means that for each individual simulation, the ratio of $dr/r$ remains constant throughout the full domain. For $[288, 576, 1152]$ zones, $dr/r = [0.016, 0.008, 0.004]$. The profiles have an initial relaxation phase of 10 dynamical times and the simulations are run to at least 100 dynamical times, which allows us to test the extent of the hydrostatic equilibrium for extended simulation times.

Radial resolution past 288 zones had little to no effect on the stability of the hydrostatic profile or our overall conclusions, yet it did make the simulations significantly more computationally heavy. Due to the minimal effect of increased resolution, we found it sufficient to run all of the simulations with a $dr/r$ value of 0.016. This ensures that the effects endured by the envelope throughout the simulations are solely due to the gravitational influence of the companion binary, and not fictitious computational effects.

\section{Cooling Testing}\label{app:cooling}

\begin{figure}[t]
    \centering
    \includegraphics[width=0.8\linewidth]{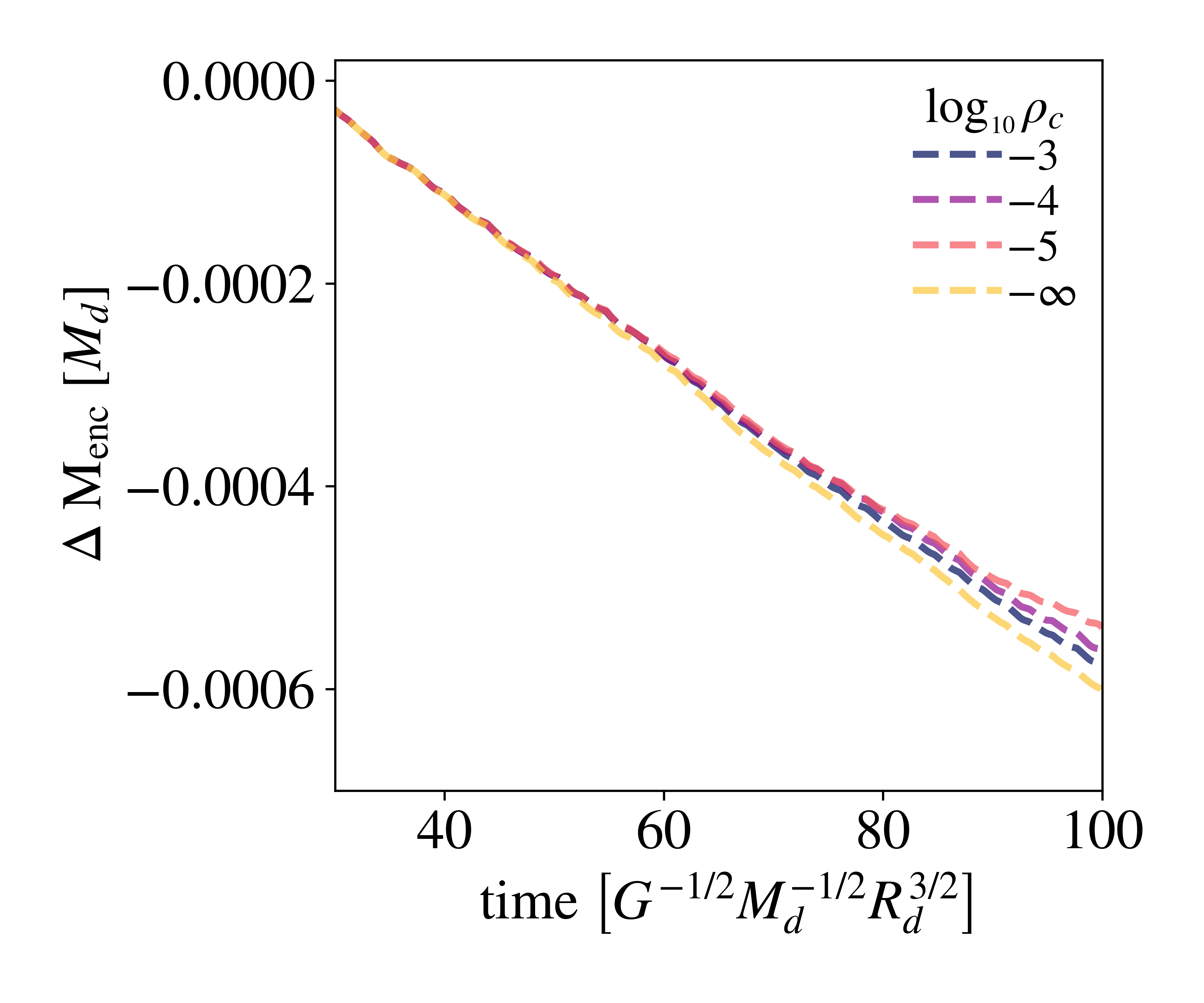}
    \caption{Change in enclosed mass over time for simulations with $\rho_c = [-3,-4,-5,-\infty]$ from darker to lighter. The value of $\rho_c = -\infty$ is essentially equivalent to the case with no cooling. }
    \label{fig:cooling_convergence}
\end{figure}

Cooling down the gas is crucial in order to form structure around the inner binary. In a realistic astrophysical context, some cooling is expected over the timescales we model due to vertical diffusion through the forming disk structures. Therefore, to capture the relevant mass transfer and accretion dynamics we find it is necessary to apply a cooling function.  In order to cool the gas, we update the pressure following the parametrized formalism from section \ref{sec:cool}. We want the overall structure of the donor to remain relatively unchanged when applying a cooling formalism. To this end, we perform a resolution study where we vary the value of the critical density, $\rho_c$, that sets the boundary between material that is cooled and material that is not. Any material with density lower than $\rho_c$ is subject to cooling. We looked at values of $\log \rho_c = [-3, -4,-5,-\infty]$ in Figure \ref{fig:cooling_convergence}. Here we plot the change in enclosed mass $\Delta M_{\rm{enc}} = M_{\rm{enc}} (t) -  M_{\rm{enc},0}$ at a distance $r=2R_d$ away from the donor. For larger values of $\rho_c$ (less negative exponent), more material will cool, while for smaller (more negative exponent) values, a smaller fraction of the material is subject to cooling. We choose to perform all of the simulations in this work with $\rho_c = 10^{-4} M_dR_d^{-3}$ as we conclude this is an appropriate value that allows material to cool down without artificially altering the rate of mass transfer from the donor onto the inner binary.

We note that we additionally tested a cooling parameter $\rho_c = 10^{-6} M_dR_d^{-3}$ but this choice showed a significant divergence in the enclosed mass, hinting at a large change in pressure in a single timestep.

\end{appendix}

\bibliography{binaries.bib,bib.bib,cooling.bib}
\bibliographystyle{aasjournalv7.1}

\end{document}